\documentclass{aa}  

\usepackage{graphicx}
\usepackage{txfonts}
\usepackage{natbib}
\usepackage{url}

\newcommand{\mygi}{MyGIsFOS}
\newcommand{\Teff}{\ensuremath{T_\mathrm{eff}}}
\newcommand{\g}{\ensuremath{g}}
\newcommand{\gf}{\ensuremath{gf}}
\newcommand{\loggf}{\ensuremath{\log\gf}}
\newcommand{\glog}{\ensuremath{\log\g}}

\newcommand{\puncms}{\mbox{\rm\,cm\,s$^{-1}$}}
\newcommand{\punkms}{\mbox{\rm\,km\,s$^{-1}$}}

\newcommand{\fei}{\ion{Fe}{i}}
\newcommand{\feii}{\ion{Fe}{ii}}
\newcommand{\alphafe}{[$\alpha$/Fe]}
\newcommand{\Vturb}{V$_\mathrm{turb}$}

\newcommand{\Vhelio}{V$_\mathrm{helio}$}

\newcommand{\feh}{[Fe/H]}
\newcommand{\ncs}{\object{NGC\, 5634}}
\newcommand{\ncsd}{\object{NGC5634-2}}
\newcommand{\ncst}{\object{NGC5634-3}}
\newcommand{\nci}{\object{NGC\, 5053}}
\newcommand{\ncis}{\object{NGC5053-69}}

\usepackage{color}

\begin{document} 

   \title{Chemical abundances of giant stars in \object{NGC 5053} and \object{NGC 5634}, two globular clusters associated with the Sagittarius dwarf Spheroidal galaxy?}

   \author{L. Sbordone \inst{1,2}
          \and
          L. Monaco \inst{3,4}
          \and
          C. Moni Bidin \inst{5}
          \and
          P. Bonifacio \inst{6}
          \and
          S. Villanova \inst{7}
          \and
          M. Bellazzini \inst{8}
          \and
          R. Ibata \inst{9}
          \and
          M. Chiba \inst{10}
          \and
          D. Geisler \inst{7}
          \and
          E. Caffau \inst{6}
          \and
          S. Duffau \inst{1,2}
          }

\institute{Millennium Institute for Astrophysics, Chile
        \and
        Pontificia Universidad Cat\'olica de Chile, Vicu\~na Mackenna 4860, Macul, Santiago, Chile \\
        \email{lsbordon@astro.puc.cl}
        \and
        European Southern Observatory, Alonso de Cordova 3107, Casilla 19001, Santiago 19, Chile
        \and
        Departamento de Ciencias Fisicas, Universidad Andres Bello, Republica 220, 837-0134 Santiago, Chile 
        \and
        Instituto de Astronom\'ia, Universidad Cat\'olica del Norte, Av. Angamos 0610, Antofagasta, Chile
        \and
	GEPI, Observatoire de Paris, PSL Resarch University, CNRS, Univ Paris Diderot, Sorbonne Paris Cit\'e, Place Jules Janssen, 92195 Meudon, France
        \and
        Departamento de Astronom\'ia, Universidad de Concepci\'on, Casilla 160-C, Concepci\'on, Chile
        \and
        INAF-Osservatorio Astronomico di Bologna, Via Ranzani 1, I-40127 Bologna, Italy
        \and
        Observatoire Astronomique de Strasbourg, Universit\'e de Strasbourg, CNRS, 11 rue de l'Universit\'e, F-67000 Strasbourg, France
        \and
        Astronomical Institute, Tohoku University, Aoba-ku, Sendai 980-8578, Japan
          }
   \authorrunning{Sbordone et al.}
   \titlerunning{Abundances in \object{NGC 5053} and \object{NGC 5634}}
   \date{Received September 15, 1996; accepted March 16, 1997}

 
  \abstract
   {The tidal disruption of the Sagittarius dwarf Spheroidal galaxy (Sgr dSph) is producing the most prominent substructure in 
   the Milky Way (MW) halo, the Sagittarius Stream. Aside from field stars, it is suspected that the Sgr dSph has lost a number of globular clusters (GC).
   Many Galactic GC are thought to have  originated in the Sgr dSph. While for some candidates an origin in the Sgr dSph
   has been confirmed owing to chemical similarities, others exist whose chemical composition has never been investigated.}
   {\nci\ and \ncs\ are two of these scarcely studied Sgr dSph candidate-member clusters.
   To characterize their composition we analyzed one giant star in \nci, and two in \ncs.}
   {We analyze high-resolution and signal-to-noise
   spectra by means of the
   MyGIsFOS code, determining atmospheric parameters and abundances for up to 21 species between O and Eu. The abundances are compared
   with those of MW halo field stars, of  unassociated MW halo globulars, and of the metal-poor Sgr dSph main body population.}
   {We derive a metallicity of  [\feii/H]=$-2.26\pm$0.10 for \nci, and of [\fei/H]=$-1.99\pm0.075$ and $-1.97\pm0.076$ for the two stars in \ncs. This makes \nci\ one of the most metal-poor 
   globular clusters in the MW.
   Both clusters display an $\alpha$  enhancement similar to the one of the halo at comparable metallicity. The two stars in \ncs\ clearly display the Na-O anticorrelation widespread
   among MW globulars. Most other abundances are in good agreement with standard MW halo trends.}
   {The chemistry of the Sgr dSph main body populations is similar to that of the halo at low metallicity. 
   It is thus difficult to discriminate between an origin of \nci\ and \ncs\ in the Sgr dSph, and one in the MW. However, the abundances
   of these clusters do appear closer to that of Sgr dSph than of the halo, favoring an origin in the Sgr dSph system.
   }


   \maketitle
%

\section{Introduction}
\label{c_intro}

It is  a fundamental  prediction of  models
of galaxy formation based on the cold dark matter (CDM) scenario, 
that dark matter haloes of the size
of that of the Milky Way grow through the accretion of
smaller subsystems \citep[see e.g.][and references therein]{Font11}. 
These resemble very much the  ``protogalactic fragments'' invoked by \citet{searle78}. The
merging of minor systems is supposed to be a common event in the early stages of the galactic history, playing a
role even in the formation of the stellar disk \citep[e.g.][]{lake89,abadi03}. Despite this general agreement,
the processes governing galaxy formation still present many obscure aspects, and understanding them is one of the greatest
challenges of modern astrophysics. For example, it has been noticed that the chemical abundance patterns of
present-day dwarf spheroidal (dSph) galaxies in the Local Group are very different from those observed among stars
in the Galactic halo \citep[see][and references therein]{vladilo03,venn04}. Most noticeably, dSphs typically
show a disappearance of $\alpha$ enhancement at lower metallicity than stars in the Milky Way, which is considered  evidence of a slow,  or bursting,  star formation history.
This is at variance with the properties of the stars
belonging to the old, spheroidal Galactic component. This clearly excludes that the known dSph can
represent the typical building blocks of larger structures such as the Galactic halo \citep{geisler07}. The
observed differences are not unexpected, however, because dSphs represent a very different environment for star
formation \citep{lanfranchi03,lanfranchi04}. In any case, the observed dSphs are evolved structures that
survived merging, while the models suggest that, although accretion events take place even today, the majority of
the merging processes occurred very early in the history of our Galaxy. The chemical peculiarities of the present-day
small satellite galaxies could have appeared later in their evolution, and the genuine building blocks could
therefore have been chemically very different from what is observed today, but more similar to the resulting merged
structures. The model of \citet{Font06} implies that the satellites that formed the halo were accreted eight to nine
Gyr ago, while the presently observed satellites were accreted only four to five Gyr ago, or are still being accreted. 

The Sagittarius (Sgr dSph) galaxy is one of the most studied systems in the Local Group because  it is the nearest
known dSph \citep{monaco04}, currently merging with the Milky Way \citep{ibata94}. It thus represents a unique
opportunity to study in detail both the stellar population of a dSph and the merging process of a minor satellite
into a larger structure. 
Among Local Group galaxies the Sgr dSph is certainly exceptional, first because of  the high metallicity of its dominant population ([Fe/H]$\sim -0.5$) compared
to its relatively low luminosity ($M_V=-$13.4, \citealt{Mateo}).
While the other galaxies of the Local Group follow a well-defined
metallicity-luminosity relation, the Sgr dSph is clearly underluminous
by almost three magnitudes  with respect to this relation \citep[see figure 5 of][]{Bonifacio05}.
The chemical composition of the Sgr dSph is also very exotic because, aside from the
aforementioned underabundance of $\alpha$-elements typical of small galaxies, all the other chemical elements
studied so far  present very peculiar patterns, clearly distinct from the Milky Way \citep{sbordone07}. However,
this behavior is observed only for stars with $[\mathrm{Fe/H}]\geq -$1. No full chemical analysis has been
performed to date on field Sgr stars of lower metallicity, but the measured abundances of $\alpha$-elements
suggest that the chemical differences with the Galactic halo should be much lower for $[\mathrm{Fe/H}]\leq -$1
\citep{monaco05}. At $[\mathrm{Fe/H}]\leq -$1.5 the Sgr dSph stellar population could be chemically indistinguishable
from the halo, at variance with other dSphs in the Local Group \citep{shetrone01,shetrone03}, although even this difference tends to disappear at lower metallicities \citep{tolstoy09}.

Decades of Galactic studies have shown that crucial information about the properties and the history of a galaxy
can be unveiled by the study of its globular clusters (GCs). For many aspects they can still be approximated as
simple, coeval, and chemically homogeneous stellar populations, although it has been known for a while that this 
is not strictly true \citep[see][for a review]{gratton12}. They thus represent a snapshot of the chemical
composition of the host galaxy at the time of their formation. The family of GCs associated with the Sgr dSph today
counts five confirmed members. However eighteen more clusters have been proposed as belonging to the Sgr dSph
(see \citealt{bellazzini02, bellazzini03a,bellazzini03} and Table 1 of \citealt{law10}, hereafter L10, for a complete census). Nevertheless, the probability of a chance alignment with the
Sgr streams is not negligible, and many objects in this large list of candidates are most probably not real
members. In their recent analysis based on new models of the Sgr tidal disruption, L10
found that only fifteen of the candidates proposed in the literature have a non-negligible probability of belonging to
the Sgr dSph. \defcitealias{law10}{L10} However, calculating the expected quantity of false associations in the
sample, they proposed that only the nine GCs with high confidence levels most likely originate from the Sgr galaxy (in good quantitative agreement with the previous analysis by \citealt{bellazzini03}).
This sample of objects with very high membership probability includes all  five of the confirmed clusters (M54,
\object{Terzan 7}, \object{Terzan 8}, \object{Arp 2}, and \object{Palomar 12}), plus \object{Berkeley 29} \citep{carraro09}, \object{Whiting 1} \citep{carraro07},
\nci, and \ncs\ \citep{bellazzini03}.
{The large list of GC candidate members is particularly interesting because the estimated total luminosity of the Sgr galaxy is comparable to that of Fornax \citep{vandenBergh00,majewski03} which, with its five confirmed GCs, is known for its anomalously high GC specific frequency \citep{vandenBergh98}. Hence, if more than five GCs were confirmed members of the Sgr family, the parent dSph would be even more anomalous than Fornax, unless its total luminosity has been largely underestimated. Estimating Sgr dSph mass is, however, difficult because  the the galaxy is being tidally destroyed, and its relatively fast chemical evolution and presence of young, metal-rich populations hint at a very massive progenitor \citep[][]{bonifacio04,sbordone07,siegel07,tolstoy09,deboer14}.}

Stimulated by the results of L10, we performed a chemical analysis of \nci\ and \ncs, since no  high-resolution study of the clusters abundances exists to date.
These objects are particularly interesting  because of  their very low metallicity
\citep[{$[\mathrm{Fe/H}]\approx-2$,}][2010 web version]{harris96}, which means that they can be used to trace the early stages of the
chemical evolution of the host galaxy. In fact, NGC\,5053 could be one of the most metal-poor GCs in the Sgr dSph family, and is also regarded as one of the most metal-poor GC known in the Milky Way.
\citet{law10} associate both clusters with the primary wrap of the trailing arm of the galaxy, in a section of
the tail probably lost by the Sgr dSph between 3 and 5 Gyr ago. Their calculations, based on their Sgr dSph merging model
and the cluster position, distance, and radial velocity, indicate that \ncs\ has a very high probability of
originating from the Sgr dSph (99.6\%), while for \nci\ this value is lower but still very significant (96\%).

\section{Observations and data reduction}

\begin{figure}
   \centering
   \includegraphics[width=\hsize]{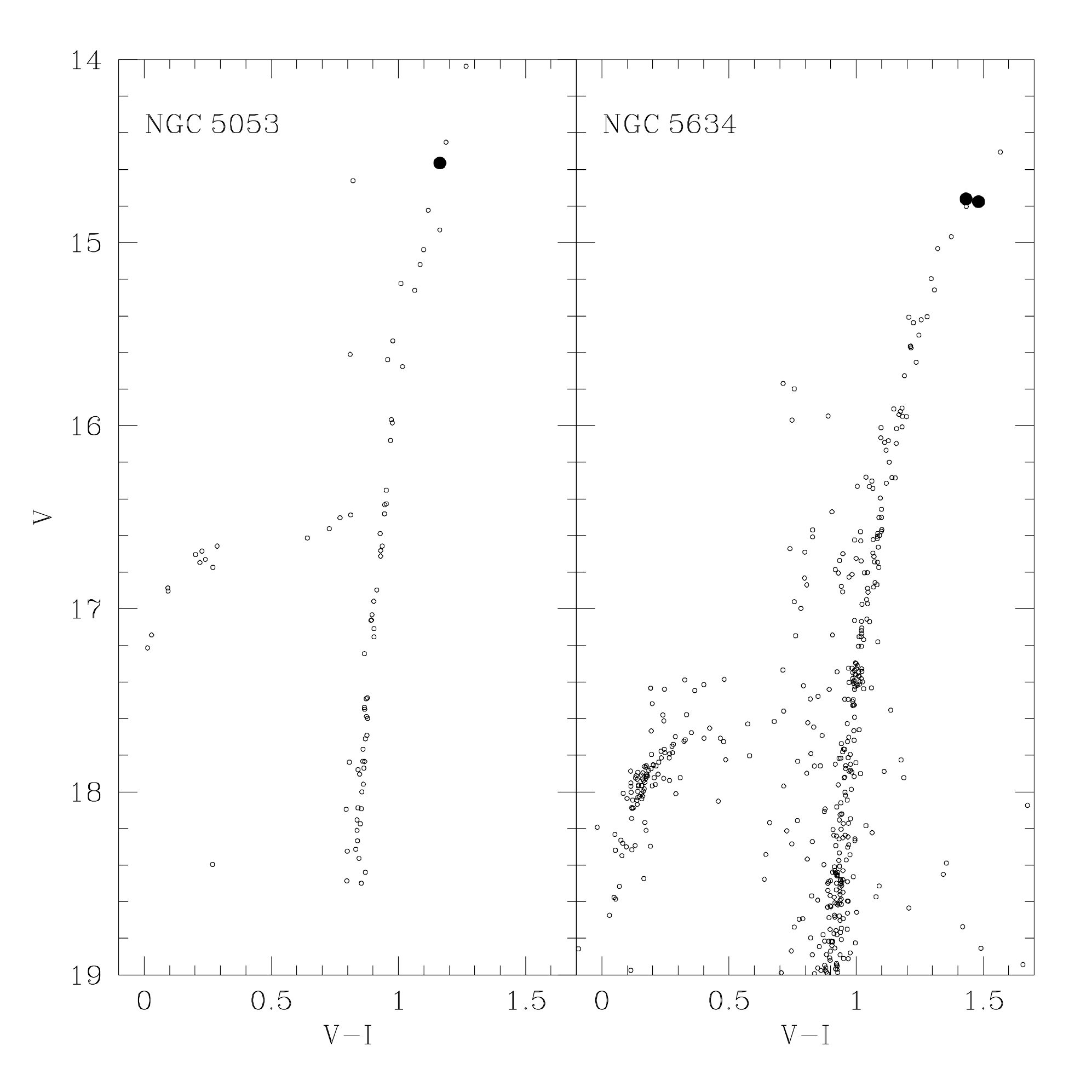}
      \caption{Color-magnitude diagrams of the target clusters (small open circles). The large full dots indicate the position of the program stars.}
         \label{f_cmd}
   \end{figure}

A high-resolution spectrum of one red giant star in \nci\ was retrieved with its calibration files
from the Keck Observatory archive\footnote{https://koa.ipac.caltech.edu/cgi-bin/KOA/nph-KOAlogin}. It was
collected with one 1800s integration with HIRES \citep{vogt94} on 2003 June 23 (program ID: C02H,
PI: I. Ivans). The spectrum covered the range 438-678 nm at a resolution of R=48\,000, resulting from
the use of a $0\farcs 86$-wide slit. The target corresponds to  object 69 in the list of standard
stars compiled by \citet{stetson00}, and its coordinates and photometric data are given in
Table~\ref{pos_param_table}. The spectrum was reduced with HIRES Redux, the IDL-based data reduction pipeline
written by Jason X. Prochaska\footnote{http://www.ucolick.org/~xavier/HIRedux/index.html}, for a typical S/N ratio { per pixel} of about 80. 
The radial velocity (RV) of the target
was measured with the {\it fxcor} IRAF\footnote{IRAF is distributed by the National Optical Astronomy
Observatories, which are operated by the Association of Universities for Research in Astronomy, Inc.,
under cooperative agreement with the National Science Foundation.} task, cross-correlating
\citep{tonry79} its spectrum with a synthetic template of a metal-poor red giant drawn from the library
of \citet{coelho05}. The observed velocity, reduced to heliocentric value was 42.2$\pm$0.7 km~s$^{-1}$.
This value matches the systemic cluster RV proposed by \citet{pryor93} and \citet{geisler95}, who measured
42.8$\pm$0.3 km~s$^{-1}$ and 42.4$\pm$1.2 km~s$^{-1}$, respectively.
{The simultaneous coincidence of the RV, metallicity (see Sect.~\ref{results_5053}), and photometry (i.e. distance,
Fig.~\ref{f_cmd}) with the cluster values confirms that the target is a cluster member.}

Two bright red giant stars in NGC\,5634 were spectroscopically observed at the SUBARU telescope on 2009
March 3 with the High Dispersion Spectrograph \citep[HDS,][]{noguchi02} in Echelle mode (program ID:
S09A-026). The targets were selected from the photometry of \citet[][hereafter B02]{bellazzini02}, and their
\defcitealias{bellazzini02}{B02}
ID numbers, coordinates,
and magnitudes are given in Table~\ref{pos_param_table}. Four exposures were collected for  star \#3 and two for
 star \#2, for a total integration time of three and two hours, respectively. The standard StdYb setup, 2x2 CCD binning, 
and the $1\farcs 2$ slit produced high-resolution spectra (R=30\,000) in the range
410-685 nm, secured on the blue and red CCDs simultaneously. Data were reduced as in \citet{monaco11}
through a combination of standard IRAF tasks, and dedicated scripts are available from the HDS
website\footnote{http://www.naoj.org/Observing/Instruments/HDS/hdsql-e.html}. 
{ The spectra of each star were then
shifted to laboratory wavelengths and merged. The final combined spectrum of NGC5634-2 had S/N$\approx$120 { per pixel}
at 600nm, while the results were of lesser quality for NGC5634-3 (S/N$\approx$80 { per pixel}) as a consequence of
shorter exposure times. }

The RV of the two stars was
measured on each extracted spectrum with the same procedure used for NGC5053-069, separately for the blue
and the red CCD. We thus obtained eight measurements for NGC5634-3 and four for NGC5634-2. For each star, velocities differ by no
more than 0.9 km~s$^{-1}$ and have a dispersion of 0.4 km~s$^{-1}$. The latter value will be assumed as the
internal error associated with our estimates, given by the average of the single measurements: $-12.8\pm$0.4
and $-20.6\pm$0.4 km~s$^{-1}$ for NGC5634-2 and NGC5634-3, respectively. 
This velocity difference is compatible with both stars being cluster members.
The velocity dispersion of this cluster is not known.
If we assume the velocities of the two stars are consistent within 1 $\sigma$ this
implies an estimate of $\sigma = 3.9$  km~s$^{-1}$ that is compatible
with what is found in several globular clusters of similar mass. The two stars could be 
compatible to less than 1 $\sigma$ and the velocity dispersion would be even higher.
The estimates of the cluster RV in the literature are scarce, and affected by large
errors. Early investigations by \citet{mayall46} and \citet{hesser86} obtained large negative values that
do not agree with our results ($-63\pm$12 and $-41\pm$9 km~s$^{-1}$, respectively). On the contrary, our
measurements agree better with \citet{peterson85}, who measured the RV of five cluster stars with an
uncertainty of 25 km~s$^{-1}$: their average value is $-$26.0 km~s$^{-1}$ with an rms of 29.1 km~s$^{-1}$,
and the resulting statistical error on the mean is 13.0 km~s$^{-1}$. Our targets lie very close on the
cluster isochrone, have identical parameters (i.e. the same distance) and metallicity, and similar RV,
hence their cluster membership is extremely likely. We conclude that the cluster RV was probably
underestimated in the literature, and the average value of $-45.1$ km~s$^{-1}$ quoted by \citet{harris96}
should be revised upward. This RV was assumed by \citetalias{law10} to assess the probability of
association with the Sgr galaxy, and it matched very well the expected value for the Sgr trailing arm. As
a consequence, the association of NGC\,5634 with this stellar stream would be less likely after the
revision of the cluster RV. However, Fig.~6 of \citetalias{law10} shows that this would still fall in the
range predicted by the model even if increased by $\sim$25 km~s$^{-1}$, hence this correction is still
compatible with their assignment.

    \begin{figure}
   \centering
   \includegraphics[width=\hsize]{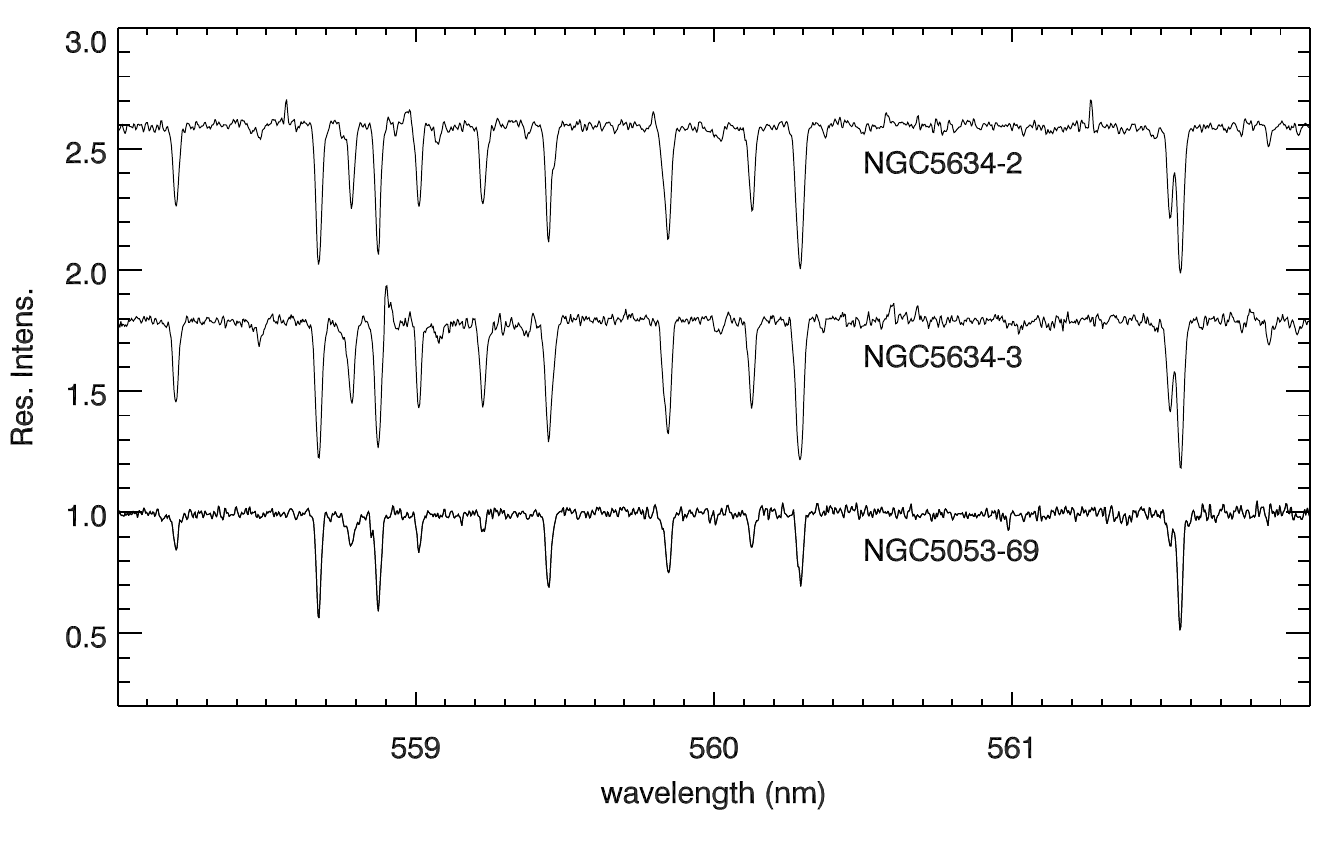}
      \caption{A sample of the spectra for the three targets around 560 nm. Spectra have been normalized and then vertically shifted for legibility.}
         \label{f_spectra}
   \end{figure}

The position of the program stars in the color-magnitude diagram (CMD) of the parent cluster is shown in
Fig.~\ref{f_cmd}, where the photometric catalogs of \citet{stetson00} and \citetalias{bellazzini02} were
used for NGC\,5053 and NGC\,6534, respectively. In Fig.~\ref{f_spectra} we show a portion of the resulting
spectra of the three stars.
Target coordinates, photometry, radial velocities, and determined atmospheric parameters are listed in Tab. \ref{pos_param_table}.

\begin{table*}
\caption{Coordinates, photometry, heliocentric radial velocities, and determined atmospheric parameters for the three targets. Photometric parameters are used for \ncis, spectroscopic ones for the two stars in \ncs\ (see Sect. \ref{choap})}             
\label{pos_param_table}           
{\centering    
\begin{tabular}{r l l c c r r c c c }     
\hline\hline       
Star              & RA               & Dec        &$V$     & $(V-I)$ & \Vhelio     & \Teff   & \glog          & \Vturb        & \feh \\
                    & (J2000)        & (J2000)  & (mag) & (mag)  & \punkms   & K       & \puncms     & \punkms    & dex \\
\hline
\object{NGC5053-69} & 13:16:35.96 & +17:41:12.8 & 14.565    & 1.163  &  42.2         & 4450    & 1.15        & 1.85       & $-$2.26\tablefootmark{a}   \\
\object{NGC5634-2}   & 14:29:30.06 & $-$05:58:39.4 & 14.776 & 1.481  & $-$12.8     & 4085    & 0.22        & 1.72       & $-$1.99                              \\
\object{NGC5634-3}   & 14:29:40.50 & $-$05:57:09.7 & 14.761 & 1.432  & $-$20.6     & 4097    & 0.45        & 1.61       & $-$1.92                              \\
\hline                    

\hline                  
\end{tabular}
}\\
\tablefoottext{a}{Using [\feii/H] for this star.}\\
\end{table*}

\section{Stellar parameters and abundance analysis}
\label{stelpar}
\subsection{Spectroscopically determined parameters}
The chemical analysis was performed by means of \mygi\footnote{\url{mygisfos.obspm.fr} will be available soon. Meanwhile contact L.S. directly} \citep[][]{sbordone14}. For this purpose, a grid of synthetic spectra covering the range between 480 nm and 690nm was computed with the following characteristics (start value, end value, step, unit): \Teff\ (4000, 5200, 200, K), \glog\ (0.5, 3, 0.5, \puncms);
\Vturb\ (1.0, 3.0, 1.0, \punkms); \feh\ ($-$4.0, $-$0.5, 0.5, dex); \alphafe ($-$0.4, 0.4, 0.4, dex). This corresponds to a grid of 3024 atmosphere models, the majority of which belonged to the MPG grid described in \citet{sbordone14}, with the exception of the ones with \Teff=4000 K and \glog=0.5, which were computed for this work. The models were computed assuming mono-dimensional, plane-parallel, and local thermodynamical equilibrium (LTE) approximations, using ATLAS 12 \citep[][]{kurucz05, sbordone04, sbordone05}. Synthetic spectra were then computed by means of SYNTHE \citep[][]{castelli05}. Atomic and molecular line data were retrieved from the R. L. Kurucz website, but the \loggf\ for the lines used in the analysis were updated according to the values provided in the second version of the Gaia-ESO (GES) ``clean'' line list. In the grid, $\alpha$ enhancement is modeled by varying in lockstep even atomic number elements between O and Ti, inclusive.

   \begin{figure}
   \centering
   \includegraphics[width=\hsize]{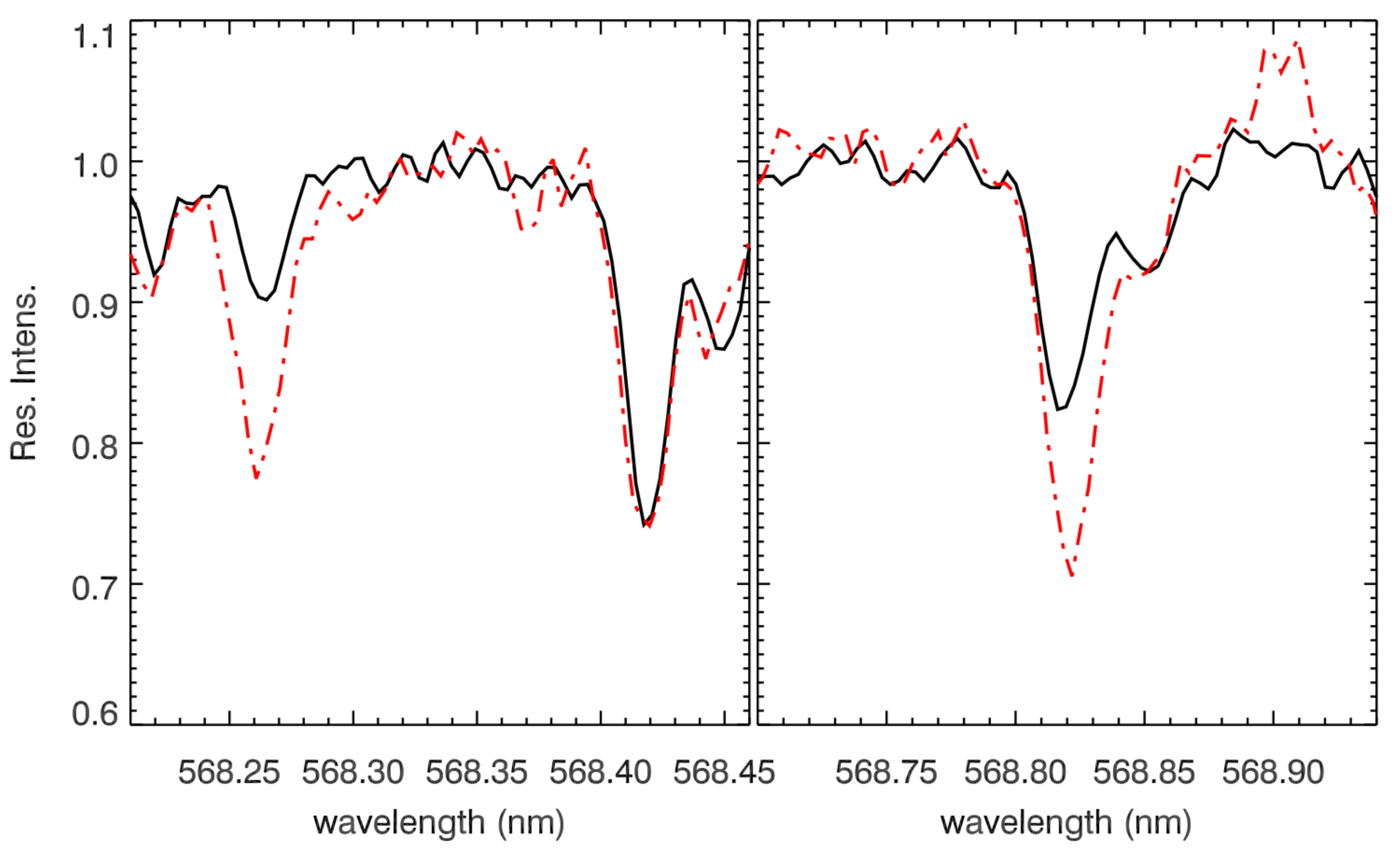}
      \caption{{ On the left, }\ion{Na}{i} 568.2 nm line and \ion{Sc}{ii} 568.4 nm lines in the spectra of stars \ncsd\ (black solid line) and \ncst\ (red dash-dotted line). { The \mygi\ fit to this specific feature is [Na/H]=-2.22 for \ncsd, and 1.72 for \ncst. On the right, the \ion{Na}{i} 568.8 nm line in the same stars. Abundances here are [Na/H]=-2.15 and -1.76.} The remarkable difference in Na abundance between the two stars is quite evident.}
         \label{na_line_fig}
   \end{figure}
 
The good quality and extensive  coverage of the spectra provided enough lines for a fully spectroscopic parameter determination. As discussed in more detail in \citet{sbordone04}, \mygi\ determines effective temperature by searching the zero of the \Teff-LEAS\footnote{Lower Energy Abundance Slope, see \citet{sbordone14}.} relation, the microturbulent velocity by eliminating the dependence of \fei\ abundance from line equivalent width, the surface gravity by imposing \fei-\feii\ ionization equilibrium. Since \alphafe\ may have a significant influence on the atmospheric structure and affect line formation, a ``global'' \alphafe\ is one of the dimensions of the synthetic grid. 
{ The synthetic grid needs to be broadened to match the combination of instrumental and macroturbulent / rotational broadening for each star. 
After checking the fit profiles on a number of unblended, metallic lines, we applied  a Gaussian broadening to the grid of FWHM=9 \punkms\ for all three stars.}
 
    \begin{figure}
   \centering
   \includegraphics[width=\hsize]{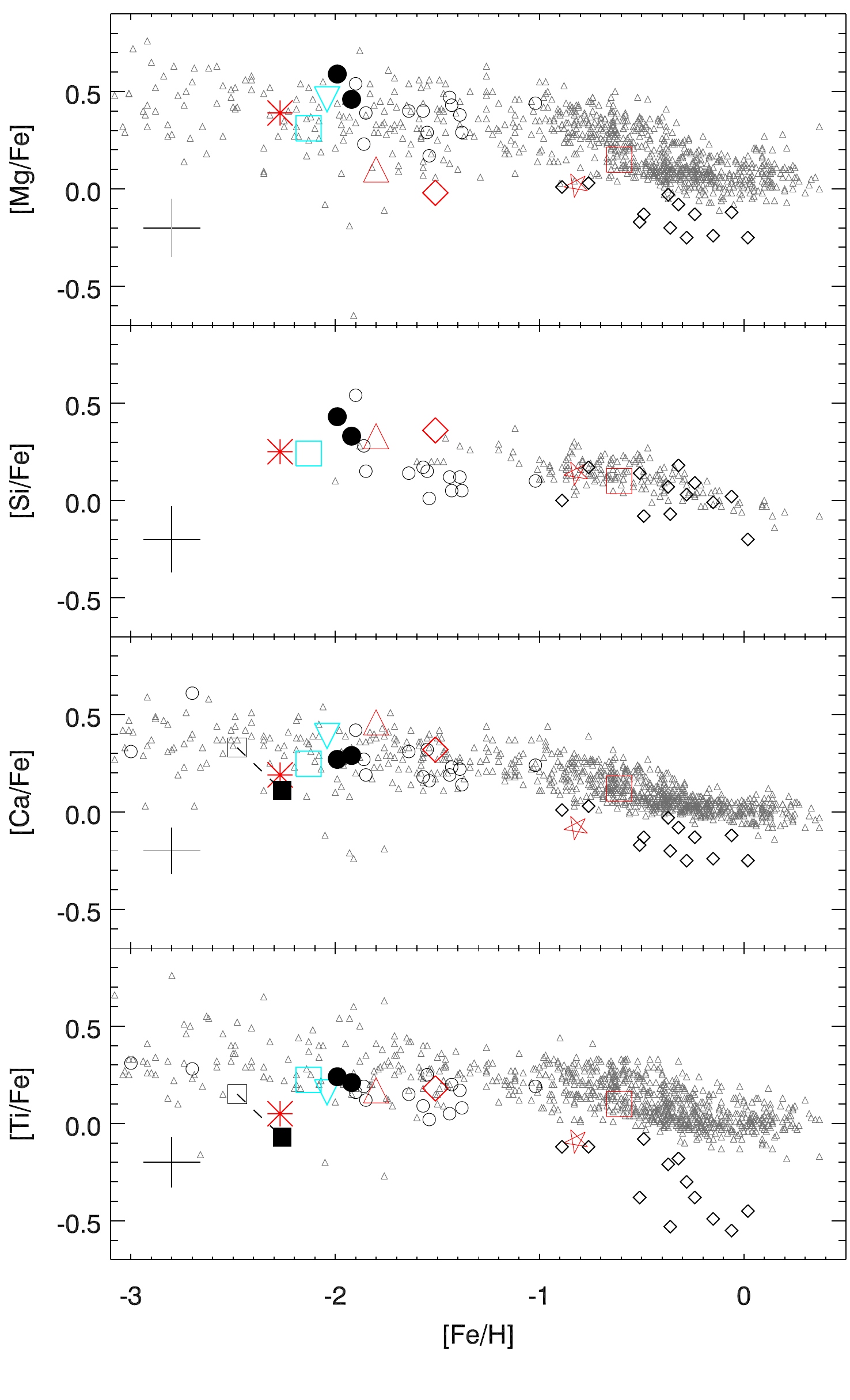}
                \caption{{ [\ion{Mg}{i}/Fe], [\ion{Si}{i}/Fe], [\ion{Ca}{i}/Fe], and [\ion{Ti}{i}/Fe] ratios plotted against [Fe/H] for  the studied stars in \nci\ (filled black squares) and \ncs\ (filled black circles), compared with Sgr dSph main body populations (open black circles, \citealt{sbordone15}; open black diamonds, \citealt{sbordone07}), and mean values for five globular clusters related to Sgr dSph: \object{Ter 8} (red asterisk), \object{Arp 2} (open red  triangle), \object{M54} (open red  diamond), \object{Pal 12} (open star), and \object{Ter 7} (open red square). Small open gray triangles are Milky Way stars \citep[][]{venn04,reddy06}. Large open cyan  symbols are MW globular clusters \object{NGC 6397} (square) and \object{NGC 5897} (triangle). The large open  black square connected by a dashed line with the filled black square represents the result for \nci\ when plotting [$\alpha$/\fei] vs. [\fei/H], with the latter represented by the open symbol. }}
         \label{alpha_fe_fig}
   \end{figure}

\begin{table*}
\caption{Detailed abundances for the three targets. N indicates the number of regions used. $\sigma$ [X/H] represents the scatter of the abundance determined from different regions, when at least two are measured for the ion.}             
\label{abund_table}      
\scriptsize
{\centering     
\begin{tabular}{llrrrrrrrrrrrrrrr}     
\hline \hline
             &             & \multicolumn{5}{l}{{\bf \object{\ncsd}}}            & \multicolumn{5}{l}{{\bf \object{\ncst}}}      & \multicolumn{5}{l}{{\bf \object{\ncis}}}             \\
Ion          & Solar       & N         & [X/H]   & $\sigma$& [X/Fe]\tablefootmark{a}     & $\sigma$ & N        &[X/H]  & $\sigma$ & [X/Fe]\tablefootmark{a} & $\sigma$   & N        &[X/H]     & $\sigma$ & [X/Fe]\tablefootmark{b} & $\sigma$       \\
             & abu.        &          &            & [X/H]   &            & [X/Fe]   &          &         & [X/H]    &        & [X/Fe]     &          &          & [X/H]    &        & [X/Fe]         \\
\hline 
\ion{O}{i}   & 8.76        & 1        & $-$1.47    & { --  }   &  0.56      & { --  }    & 1        & -1.63   & { --  }    & 0.29    & { --  }      & --       &  --      & { --  }    & --      & { --  } \\
\ion{Na}{i}  & 6.30        & 2        & $-$2.19    & { 0.05}   & $-$0.20    & { 0.09}    & 3        & $-$1.79 & { 0.09}    &  0.12   & { 0.12}      &  1       & $-$2.05  & { --  }    &  0.21   & { --  } \\
\ion{Mg}{i}  & 7.54        & 1        & $-$1.40    & { --  }   &  0.59      & { --  }    & 1        & $-$1.46 & { --  }    &  0.46   & { --  }      & --       &  --      & { --  }    & --      & { --  } \\
\ion{Al}{i}  & 6.47        & --       & --         & { --  }   & --         & { --  }    & 1        & $-$1.32 & { --  }    &  0.59   & { --  }      & --       &  --      & { --  }    & --      & { --  } \\
\ion{Si}{i}  & 7.52        & 3        & $-$1.56    & { 0.16}   &  0.43      & { 0.18}    & 3        & $-$1.58 & { 0.16}    &  0.33   & { 0.17}      & --       &  --      & { --  }    & --      & { --  } \\
\ion{Ca}{i}  & 6.33        & 12       & $-$1.73    & { 0.11}   &  0.27      & { 0.13}    & 9        & $-$1.62 & { 0.08}    &  0.29   & { 0.11}      &  9       & $-$2.15  & { 0.08}    &  0.10   & { 0.13} \\
\ion{Sc}{ii} & 3.10        & 10       & $-$1.78    & { 0.09}   &  0.25      & { 0.18}    & 8        & $-$1.65 & { 0.06}    &  0.27   & { 0.16}      &  9       & $-$2.09  & { 0.13}    &  0.10   & { 0.17} \\
\ion{Ti}{i}  & 4.90        & 10       & $-$1.75    & { 0.05}   &  0.24      & { 0.09}    & 9        & $-$1.70 & { 0.10}    &  0.21   & { 0.13}      &  3       & $-$2.33  & { 0.13}    & $-$0.07 & { 0.17} \\
\ion{Ti}{ii} & 4.90        & 2        & $-$1.71    & { 0.09}   &  0.32      & { 0.18}    & 2        & $-$1.54 & { 0.13}    &  0.38   & { 0.20}      &  3       & $-$2.08  & { 0.10}    &  0.10   & { 0.14} \\
\ion{V}{i}   & 4.00        & 16       & $-$2.19    & { 0.12}   & $-$0.20    & { 0.14}    & 16       & $-$2.09 & { 0.11}    & $-$0.18 & { 0.13}      & --       &  --      & { --  }    & --      & { --  } \\
\ion{Cr}{i}  & 5.64        & 3        & $-$2.16    & { 0.05}   & $-$0.16    & { 0.09}    & 2        & $-$2.09 & { 0.16}    & $-$0.17 & { 0.17}      &  2       & $-$2.75  & { 0.05}    & $-$0.49 & { 0.12} \\
\ion{Mn}{i}  & 5.37        & 5        & $-$2.36    & { 0.15}   & $-$0.37    & { 0.16}    & 5        & $-$2.31 & { 0.14}    & $-$0.40 & { 0.16}      &  1       & $-$2.88  & { --  }    & $-$0.62 & { --  } \\
\ion{Fe}{i}  & 7.52        & 52       & $-$1.99    & { 0.07}   &  --        & { 0.11}    & 55       & $-$1.92 & { 0.08}    &  --     & { 0.11}      & 27       & $-$2.48  & { 0.08}    & $-$0.22 & { 0.13} \\
\ion{Fe}{ii} & 7.52        & 9        & $-$2.03    & { 0.16}   &  --        & { 0.22}    & 7        & $-$1.92 & { 0.15}    &  --     & { 0.22}      &  5       & $-$2.26  & { 0.10}    & --      & { 0.14} \\
\ion{Co}{i}  & 4.92        & 4        & $-$1.94    & { 0.07}   &  0.05      & { 0.10}    & 3        & $-$1.94 & { 0.07}    & $-$0.02 & { 0.10}      & --       &  --      & { --  }    & --      & { --  } \\
\ion{Ni}{i}  & 6.23        & 7        & $-$2.16    & { 0.15}   & $-$0.17    & { 0.17}    & 6        & $-$2.03 & { 0.15}    & $-$0.11 & { 0.17}      &  1       & $-$2.45  & { --  }    & $-$0.19 & { --  } \\
\ion{Cu}{i}  & 4.21        & --       & --         & { --  }   & --         & { --  }    & 1        & $-$2.43 & { --  }    & $-$0.52 & { --  }      & --       &  --      & { --  }    & --      & { --  } \\
\ion{Zn}{i}  & 4.62        & 1        & $-$2.06    & { --  }   & $-$0.07    & { --  }    & 1        & $-$2.15 & { --  }    & $-$0.23 & { --  }      &  1       & $-$2.42  & { --  }    & $-$0.16 & { --  } \\
\ion{Y}{ii}  & 2.21        & 4        & $-$2.43    & { 0.14}   & $-$0.40    & { 0.21}    & 4        & $-$2.25 & { 0.13}    & $-$0.33 & { 0.20}      &  2       & $-$2.89  & { 0.10}    & $-$0.63 & { 0.14} \\
\ion{Ba}{ii} & 2.17        & 1        & $-$2.12    & { --  }   & $-$0.09    & { --  }    & 1        & $-$1.82 & { --  }    &  0.10   & { --  }      &  3       & $-$2.51  & { 0.07}    & $-$0.25 & { 0.12} \\
\ion{Eu}{ii} & 0.52        & 1        & $-$1.37    & { --  }   &  0.66      & { --  }    & 1        & $-$1.17 & { --  }    &  0.75   & { --  }      & --       &  --      & { --  }    & --      & { --  } \\
\hline                  
\end{tabular}
}\\
\tablefoottext{a}{[X/Fe] computed against \fei\ for neutral species except \ion{O}{i}, against \feii\ for ionized species and  \ion{O}{i}.}\\
\tablefoottext{b}{All [X/Fe] computed against \feii.}
\end{table*}

\subsection{Photometric parameters estimates}

In addition to the fully spectroscopic parameter estimates, we also estimated \Teff\ from the de-reddened color $(V-I)$ through the equations of
\citet{alonso99,alonso01}, assuming $E(V-I)$=0.08~mag for \ncs\ \citep{bellazzini02}, and $E(B-V)$=0.017~mag for \nci\ 
\citep{nemec04}, with $E(V-I)=1.34\times E(B-V)$ \citep{dean78}. The adopted color--temperature relation is independent
of metallicity.

The surface gravity was calculated through the equation
$$
\log{\mathrm{g}}=\log{\mathrm{M}}+4\log{\mathrm{T_{eff}}}+0.4(M_\mathrm{V}+BC)-12.503,
\label{e_logg}
$$
obtained from basic relations, where $M$ is the mass, BC is the bolometric correction, and the solar values $T_{\sun}$=5777 K,
$\log{g_{\sun}}$=4.44, $M_\mathrm{bol,\sun}$=4.75 are assumed in the calculation of the constant term. The stellar mass was fixed to
0.80$\pm$0.05 M$_{\sun}$ for all the stars, and BC was deduced from the temperature interpolating the tables of \citet{alonso99}.
The absolute magnitude in the $V$ band was estimated from the cluster reddening, defined as before, and distance modulus
$(m-M)_V$=16.23 \citep{harris96} and 17.36~magnitudes \citep{bellazzini02} for \nci\ and \ncs, respectively. Starting from these values for \Teff\ and \glog, we re-derived \Vturb\ and abundances as we did for the fully spectroscopic case: for brevity, we will refer to this set of parameters as ``photometric'' parameters from now on.

The temperature of the two stars in \ncs\ are very similar to the spectroscopic values ($\Delta T_\mathrm{eff}<40$~K), and the gravities are higher but still compatible within uncertainties. The photometric gravity of \ncis\ is, on the contrary, 0.9~dex higher than the spectroscopic value. In particular, the spectroscopic parameters of this star cannot satisfy the basic relation above, unless one or more of the other input quantities (cluster parameters, stellar mass, BC) are revised to unrealistic values.

\subsection{Choice of adopted atmospheric parameters}
\label{choap}

The discrepancy between photometric and spectroscopic atmospheric parameters is a well-documented fact for giant stars around or below [Fe/H]=$-$2. In two recent studies, for instance, \citet{mashonkina11} and \citet{bergemann12} study the case of the metal-poor giant \object{HD 122563}, whose parameters can be reliably derived from photometry and Hipparcos-based parallaxes. Said parameters (\Teff=4665 K, \glog=1.64, \Vturb=1.61 \punkms, \feh=-2.60) are quite close to the ones photometrically derived for \ncis, and, when using them in a 1D-LTE analysis, \object{HD 122563} shows a \fei--\feii\ imbalance and residual LEAS quite close to the ones \ncis\ displays in the present study. \citet{mashonkina11} manages to recover a satisfactory ionization balance for \object{HD 122563} when treating \fei\ line formation taking into account departures from LTE  (NLTE), but is left with an unacceptable LEAS, albeit a reduced one with respect to  the LTE case. \citet{bergemann12} also employs NLTE line formation for \fei, but in association with horizontally averaged 3D hydrodynamical models, which preserve the vertical temperature structure of 3D models. Despite this further refinement, the LEAS problem of \object{HD 122563} remains unsolved, likely indicating that the horizontal temperature variations (that are averaged out in the \citealt{bergemann12} models) must be accounted for {\em together} with NLTE to properly describe \fei\ line formation in these cool, metal-poor  giants. It is thus evident that NLTE affects \fei\ line formation in a relevant way in stars like \ncis, and likely, to a lower extent, \ncsd\ and \ncst. 
   
In the case of \ncis, the LTE treatment significantly underestimates Fe ionization and upper-level populations \citep{mashonkina11}, leading to an overestimate of \fei\ line strength, and a consequent underestimate of the derived abundance. We thus decided to adopt the photometric set of parameters for \ncis, employ \feii\ for the determination of \feh, and derive all the [X/Fe] ratios with respect to [\feii/H]. 

The situation is less straightforward in the case of the two stars in \ncs. The two stars show quasi-identical spectra, which  the spectroscopic parameter estimate underscores by providing practically equal parameters. Despite this agreement, they show a significant $(V-I)$ difference, which translates to a 86 K photometric \Teff\ difference. Most likely, this is a consequence of \ncst\ being a second-generation star, as clearly indicated by its Na abundance (see Sect. \ref{results} and Fig. \ref{na_line_fig}). The photometric effect of abundance variations in globular clusters was studied in detail in \citet{sbordone11}, where it can be seen that second-generation stars in the upper RGB become slightly bluer than first-generation ones, essentially because of the flux lost in the UV due to stronger NH absorption being transferred more in the $V$ band than in the $I$. An inspection of \citet{sbordone11} ``reference'' and ``CNONa2'' isochrones around the spectroscopic temperature of \ncst\ shows an effect of about 0.04 magnitudes in $(V-I)$, strikingly similar to the difference in color between \ncsd\ and \ncst. This cannot be taken strictly at face value: the \citet{sbordone11} calculations were performed at a slightly higher metallicity ([Fe/H]=-1.62), and the assumed variations of C, N, and O abundances were likely larger than the ones encountered in \ncs. However, it is a clear indication that, in the absence of calibrations computed taking into account abundance variations such as the one encountered in globular clusters, photometric temperature estimates might be significantly skewed in second-generation stars { in the upper part of the RGB}.
It is interesting to note how star \ncsd, on the other hand, shows an excellent agreement between photometric and spectroscopic temperature, and a photometric gravity that is only 0.33 dex higher than the spectroscopic. { Although we are considering one single star here, this hints at the fact that NLTE effects on \fei\ on \Teff\ and gravity are strongly reduced at the metallicity of \ncs}. For this reason, we feel confident that spectroscopic parameters can be employed in this case, and we consider them preferable owing to their insensitivity to the effects of CNO abundance variations.

   \begin{figure}
   \centering
   \includegraphics[width=\hsize]{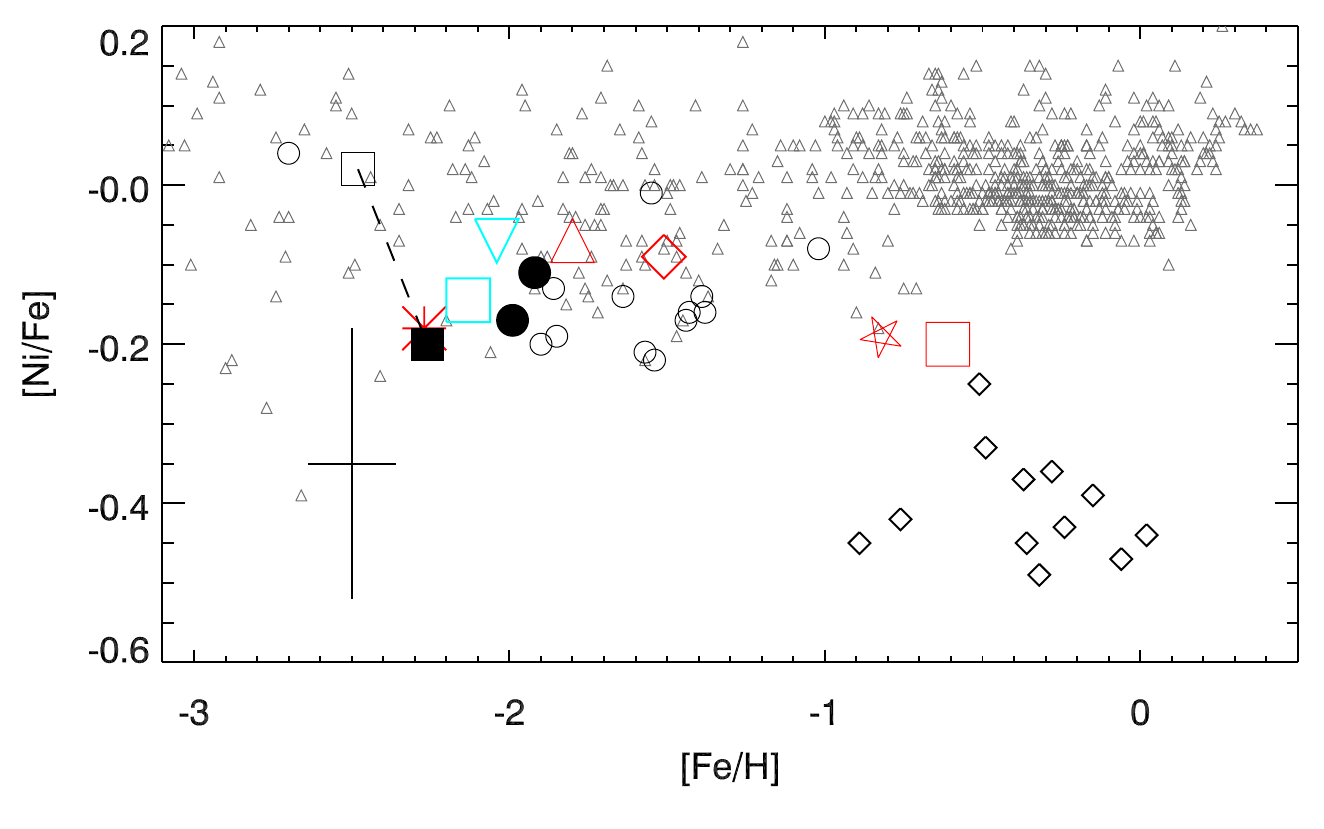}
      \caption{[Ni/Fe] vs. \feh\ for the same samples, and using the same symbols as in Fig. \ref{alpha_fe_fig}.}
         \label{ni_fe_fig}
   \end{figure}

In Tab. \ref{alternate_param_table} we list the variations in [X/H] in each star that would have originated from employing the ``rejected'' set of parameters for the star, i.e. the photometric parameters for the two stars in \ncs, and the spectroscopic ones in \ncis. { Here, the strongest variations are, as expected, encountered in the ionized species in \ncis. Since the rejected parameter set would force gravity to recover \fei--\feii\ ionization equilibrium, abundances for \feii\ and other ionized species decrease consistently by 0.25-0.35 dex. Most neutral species decrease of about 0.05 dex (likely due to the roughly 100 K lower \Teff\ in the rejected set), with the exception of Zn, which shows a stronger effect. The effect is the opposite in \ncsd, where employing the rejected photometric set would break Fe ionization equilibrium, raising the abundance of ionized species, although, in this case, \feii\ varies more than most of the other ionized elements. Also affected is \ion{O}{i}, since the [\ion{O}{i}] 630 nm line is, naturally, sensitive to pressure. Finally, \ncst\ has very close photometric and spectroscopic parameters, leading to almost identical abundances. As usually observed in these cases, [X/Fe] abundance ratios produced taking care of matching ionization stages are quite robust against variations in atmospheric parameters.}

\begin{table}
\caption{Variation in [X/H] when using for each star the alternate parameter set with respect to the one chosen. Variations are computed as rejected -- chosen.}             
\label{alternate_param_table}      
\centering     
\footnotesize
\begin{tabular}{lrrr}
\hline\hline
                                                &{\bf \ncis}  &{\bf \ncsd}    &{\bf \ncst}    \\    
                                                &\Teff=4343   &\Teff=4071     &\Teff=4135     \\  
                                                &\glog=0.26   &\glog=0.55     &\glog=0.55     \\
                                                &\Vturb=1.7   &\Vturb=1.7     &\Vturb=1.6     \\ 
                                                &$\Delta$[X/H] &$\Delta$[X/H]  &$\Delta$[X/H] \\
\hline
\ion{O}{i}  &  --          & 0.19         & 0.04            \\  
\ion{Na}{i} & $-$0.01      & $-$0.05      & 0.02            \\         
\ion{Mg}{i} & --           & $-$0.01      & 0.00            \\  
\ion{Al}{i} & --           & --           & 0.02            \\     
\ion{Si}{i} & --           & 0.06         & 0.01            \\  
\ion{Ca}{i} &  0.02        & $-$0.06      & $-$0.01         \\   
\ion{Sc}{ii}& $-$0.37      & 0.16         & 0.03            \\   
\ion{Ti}{i} & $-$0.07      & $-$0.05      & 0.04          \\   
\ion{Ti}{ii}& $-$0.26      & 0.07         & $-$0.02         \\   
\ion{V}{i}  & --           & 0.04         & 0.08          \\    
\ion{Cr}{i} & $-$0.03      & $-$0.05      & 0.05          \\    
\ion{Mn}{i} &  0.01        & $-$0.04      & 0.05          \\      
\ion{Fe}{i} & $-$0.04      & $-$0.02      & 0.03          \\      
\ion{Fe}{ii}& $-$0.27      & 0.21         & 0.06          \\       
\ion{Co}{i} & --           & 0.02         & 0.05          \\   
\ion{Ni}{i} & $-$0.07      & 0.00         & 0.04          \\      
\ion{Cu}{i} & --           & --           & 0.04          \\          
\ion{Zn}{i} & $-$0.14      & 0.10         & $-$0.02         \\         
\ion{Y}{ii} & $-$0.29      & 0.07         & 0.02            \\      
\ion{Ba}{ii}& $-$0.26      & 0.12         & 0.04            \\         
\ion{Eu}{ii}& --           & 0.17         & 0.04            \\       
\hline 
\end{tabular}
\end{table}

\subsection{Detailed abundances}

{ After parameter determination, \mygi\ derived detailed abundances for all the elements for which regions had been provided and viable lines were found. All the employed regions, as well as the observed and best-fitting synthetic profiles, can be obtained online via Vizier (see Appendix \ref{mygi_appendix}). Hyper fine splitting (HFS) values for the used lines were instead derived from the fourth version of the GES line list \citep[][]{heiterprep}, and used for the lines of \ion{Sc}{ii}, \ion{V}{i}, \ion{Mn}{i}, \ion{Co}{i}, \ion{Cu}{i}, \ion{Ba}{ii}, and \ion{Eu}{ii}. Isotopic mixture for Eu was derived from \citet{anders89}. 

Abundances for all elements were measured within \mygi, with the exception of oxygen, whose abundance was determined in \ncs\ by measuring the [\ion{O}{i}] 630.0304 nm line}. The line could be measured by MyGIsFOS in \ncsd, but not in \ncst\  since the line was contaminated by telluric absorptions. In the latter case, the region around the [\ion{O}{i}] 630.0304 nm line was decontaminated from telluric lines according to the following procedure. First of all, we calculated a synthetic telluric spectrum where parameters were varied in order  to match the depth and width of the unblended telluric features. Then we { divided the spectral region surrounding the [\ion{O}{i}] 630.0304 nm line by this spectrum}. We then proceeded to fit the oxygen line in both stars as follows: for each star, an ATLAS 12 model at the final parameters and abundances was computed, from which two small grids of synthetic spectra were computed in the region surrounding the oxygen line at varying O abundances. The {\tt fitprofile} line-fitting code \citep{thygesen15} was then employed to derive the  best-fitting O abundance by $\chi^2$ minimization. { For consistency, {\tt fitprofile} was employed to derive O abundances in both \ncsd\ and \ncst}.

\section{Results}
\label{results}

Derived abundances are in Tab. \ref{abund_table}. The assumed solar abundances used in the grid are also listed in this table, and were taken from the compilation of \citet{lodders09}, except for O, Fe, and Eu, which where taken from \citet{caffau11}, while Tab. \ref{abund_variation_table} lists the impact on the derived abundances for star \ncst\ of altering each atmospheric parameter by an amount roughly equivalent to the estimated error.

In Figs. \ref{alpha_fe_fig} to \ref{nieu_fig}  we plot various chemical abundance ratios in the three studied stars compared with values in different Sgr dSph populations, other globular clusters, and stars in the Milky Way disk and halo. { In all these figures, typical error bars for the present analysis are overplotted: they correspond to line-to-line scatter as indicated in Tab. \ref{abund_table}. Where only one line was measured, we show a default 0.15 dex error bar for [X/Fe], and the corresponding bar is traced in gray}. For the globular clusters associated with Sgr dSph (red symbols), average cluster values have been retrieved from \citet{carretta14}, and come from \citet{carretta14} and \citet{mottini08} for \object{Terzan 8}, \citet{mottini08} for \object{Arp 2}, \citet{carretta10} for \object{M 54}, \citet{cohen04} for \object{Palomar 12}, and \citet{sbordone05b} for \object{Terzan 7}. Milky Way stars (gray symbols) are taken from the \citet{venn04} compilation, while  the two MW comparison clusters \object{NGC 6397} and \object{NGC 5897} (cyan symbols) come from \citet{lind11} and \citet{koch14}, respectively.
The data points referring to the Sgr dSph metal-poor main body populations come from a companion work to the present one, based on UVES \citep[][]{dekker00} spectra. The full analysis of these stars will be presented in \citet{sbordone15}, here we just point out that these stars are similar (albeit slightly less evolved) to the ones studied here, and have been analyzed by means of \mygi, using the same grid and the same region list employed in this paper. Metal-rich (\feh$>$-1.0) Sgr dSph data points are taken from \citet{sbordone07}.

\subsection{\ncs}
\label{results_5634}
The two stars examined are very high-luminosity red giants, as made evident by their very low gravity, which required a slight extrapolation of the \mygi\ grid in the gravity dimension (Table \ref{pos_param_table}). In fact, both stars show a clear emission component on H$\alpha$ wings (see Fig. \ref{5634_halpha_fig}), apparently asymmetric with respect to the absorption line, which could be interpreted as evidence of mass loss from the star. The emission appears to be more blueshifted (or more blue-asymmetric) in star \ncst, leading to a slightly different center of the absorption component. 

The two stars are extremely similar, with atmospheric parameters whose differences are well within the observational uncertainties. Chemical abundances are also, in most cases, extremely similar between the two stars. The \feh\ ($-1.94\pm$0.08 and $-1.93\pm$0.08
 for \ncst) indicate a cluster metallicity in excellent agreement with existing photometric estimates \citep[\feh=$-$1.94, see][and references therein]{bellazzini02}. 

\begin{table*}
\caption{Variations in the derived abundances due to varying atmosphere parameters in star \ncst. { Since the parameters of the three stars are quite similar, these systematic errors can be applied to all  three stars.}}             
\label{abund_variation_table}      
\centering     
\footnotesize
\begin{tabular}{lrrrrrrrrrrrr}  
\hline\hline
Variation               & \ion{O}{i} & \ion{Na}{i} & \ion{Mg}{i} & \ion{Al}{i} & \ion{Si}{i} & \ion{Ca}{i} & \ion{Sc}{i} & \ion{Ti}{i} & \ion{Ti}{ii} & \ion{V}{i} & \ion{Cr}{i}\\   
\hline
\Teff\ $+$50 K          & 0.00 & 0.04 &  0.03 &  0.03 & $-$0.01 &  0.08 & $-$0.01 &  0.09 &  0.00 &  0.10 &  0.09 \\    
\Teff\ $-$50 K          & $-$0.01&$-$0.04 & $-$0.04 & $-$0.03 &  0.00 & $-$0.06 & $-$0.01 & $-$0.09 &  0.00 & $-$0.10 & $-$0.09 \\
\glog\ $+$0.3           & 0.10 &$-$0.03 & $-$0.01 &  0.01 &  0.05 & $-$0.03 &  0.13 & $-$0.02 &  0.06 & $-$0.01 & $-$0.02 \\
\glog\ $-$0.3           & $-$0.12 & 0.03 &  0.01 & $-$0.01 & $-$0.05 &  0.05 & $-$0.14 &  0.02 & $-$0.05 &  0.01 &  0.02 \\
\Vturb\ $+$0.2 \punkms  & $-$0.02&$-$0.02 & $-$0.03 & $-$0.01 & $-$0.01 & $-$0.07 & $-$0.06 & $-$0.03 & $-$0.07 & $-$0.01 & $-$0.08 \\
\Vturb\ $-$0.2 \punkms  & 0.00 & 0.02 &  0.03 &  0.01 &  0.01 &  0.11 &  0.06 &  0.07 &  0.08 &  0.01 &  0.11 \\
\alphafe\ $+$0.2        & 0.01&$-$0.01 &  0.00 & $-$0.01 &  0.04 &  0.02 &  0.04 &  0.05 &  0.05 &  0.00 &  0.00 \\
\alphafe\ $-$0.2        & $-$0.07& 0.02 & $-$0.01 &  0.01 & $-$0.04 & $-$0.03 & $-$0.05 & $-$0.05 & $-$0.05 &  0.00 &  0.00 \\
\\
\hline
Variation               &\ion{Mn}{i} & \ion{Fe}{i} & \ion{Fe}{ii} & \ion{Co}{i} & \ion{Ni}{i} & \ion{Cu}{i} & \ion{Zn}{i} & \ion{Y}{ii} & \ion{Ba}{ii} & \ion{Eu}{ii} & & \\
\hline
\Teff\ $+$50 K          &  0.07 &  0.04 & $-$0.08 &  0.06 &  0.05 &  0.07 & $-$0.04 &  0.01 &  0.02 & $-$0.02 &\\
\Teff\ $-$50 K          & $-$0.07 & $-$0.04 &  0.08 & $-$0.06 & $-$0.05 & $-$0.08 &  0.04 & $-$0.02 & $-$0.02 &  0.02 & \\
\glog\ $+$0.3           & $-$0.01 &  0.00 &  0.19 &  0.03 &  0.02 & $-$0.03 &  0.07 &  0.08 &  0.14 &  0.19 & \\
\glog\ $-$0.3           &  0.01 &  0.00 & $-$0.16 & $-$0.03 & $-$0.02 &  0.02 & $-$0.07 & $-$0.07 & $-$0.13 & $-$0.15 & \\
\Vturb\ $+$0.2 \punkms  & $-$0.03 & $-$0.06 & $-$0.06 & $-$0.01 & $-$0.03 & $-$0.04 & $-$0.05 & $-$0.08 & $-$0.23 & $-$0.01 & \\
\Vturb\ $-$0.2 \punkms  &  0.03 &  0.07 &  0.06 &  0.01 &  0.04 &  0.05 &  0.06 &  0.09 &  0.30 &  0.01 & \\
\alphafe\ $+$0.2        &  0.00 &  0.00 &  0.07 &  0.01 &  0.00 & $-$0.02 &  0.10 &  0.03 &  0.06 &  0.10 & \\
\alphafe\ $-$0.2        &  0.00 & $-$0.01 & $-$0.07 & $-$0.01 &  0.00 &  0.02 & $-$0.09 & $-$0.03 & $-$0.06 & $-$0.09 & \\           
\hline
\end{tabular}
\end{table*}

Oxygen appears enhanced with respect to the solar ratio in a fashion compatible with typical halo values at this metallicity in \ncsd, less so in \ncst. The  [O/Fe] ratio differs by 0.27 dex between the two stars, anticorrelating with the difference in Na abundance.

Sodium is the element that displays the most strikingly different abundance between the two stars. They differ by 0.34 dex, well beyond the line-to-line dispersion, with star \ncst\ being the more Na-rich. The difference is readily visible in the spectrum, as shown in Fig. \ref{na_line_fig}, where \ion{Na}{i} features appear much stronger in \ncst, despite almost identical parameters. The clear Na abundance difference, together with the opposite O abundance difference, make it highly probable that \ncs\ displays a significant Na abundance spread, and possibly a Na-O abundance anticorrelation, as is almost universally observed in globular clusters. It should be remarked that the observed Na abundance difference is about half of the typical full extent of the Na abundance spread as observed in most GCs \citep[see for instance Fig. 2 in ][]{gratton12}. A sample of only two stars, obviously, does not allow one to infer the full extent of the spread, nor the numerical significance of the second generation in \ncs. 

NLTE corrections are available for two of the \ion{Na}{i} lines we used (568.2nm and 568.8nm) as derived from the calculations of \citet{lind11b}, and made available on the web through the {\tt INSPECT}\footnote{\url{http://www.inspect-stars.com}} interface. The available calculations have a lower \glog\ limit of 1.0, so we could not test the values for the atmospheric parameters of our stars. However, corrections appear to be very small in this parameter domain (0.03 dex for the 568.2nm line, and 0.05 for the 568.8nm line with the parameters and line strength of star \ncsd).

\begin{figure}
   \centering
   \includegraphics[width=\hsize]{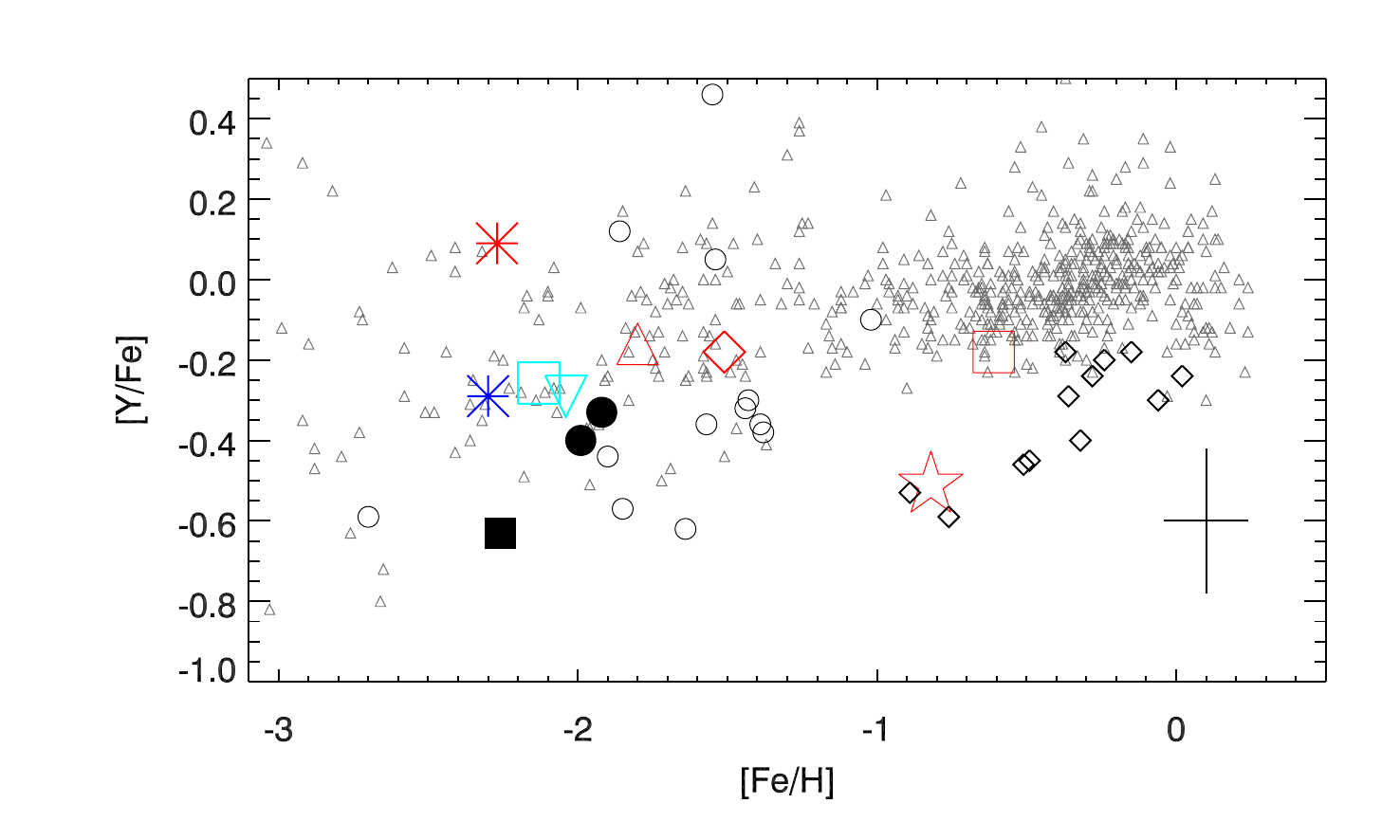}
      \caption{[Y/Fe] plotted against [Fe/H]. Symbols are the same as in  Fig. \ref{alpha_fe_fig} except for the large blue asterisk, which  represents the average Y abundance in \object{Ter 8} as measured by \citet{mottini08}. Here, in Fig. \ref{ba_fe_fig}, and in Fig. \ref{bay_fe_fig} only the filled point appears for \nci\, since both Ba and Y are always compared to \feii.}
         \label{y_fe_fig}
   \end{figure}

\begin{figure}
   \centering
   \includegraphics[width=\hsize]{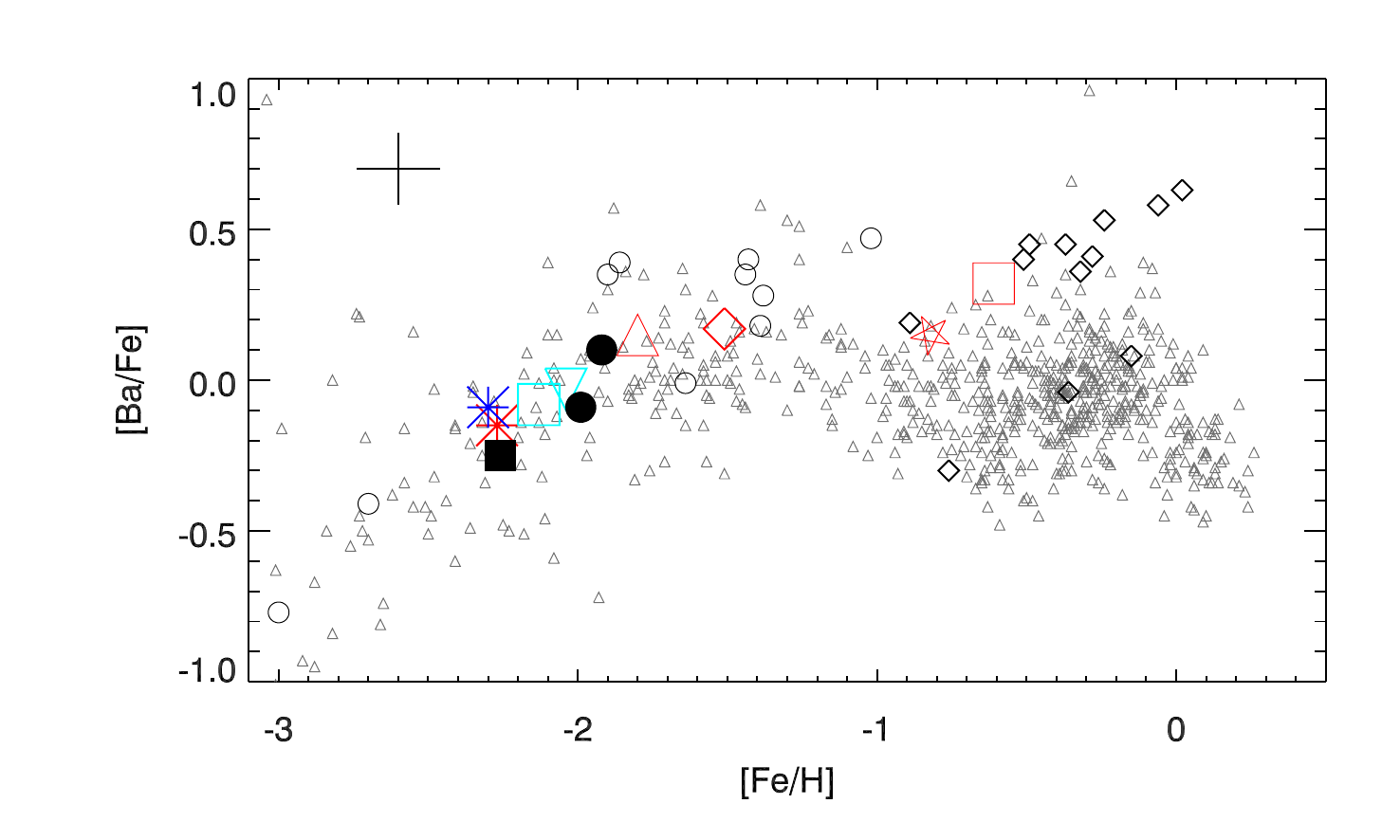}
      \caption{{ [Ba/Fe] plotted against [Fe/H], symbols as in  Fig. \ref{y_fe_fig}.}}
         \label{ba_fe_fig}
   \end{figure}

\begin{figure}
   \centering
   \includegraphics[width=\hsize]{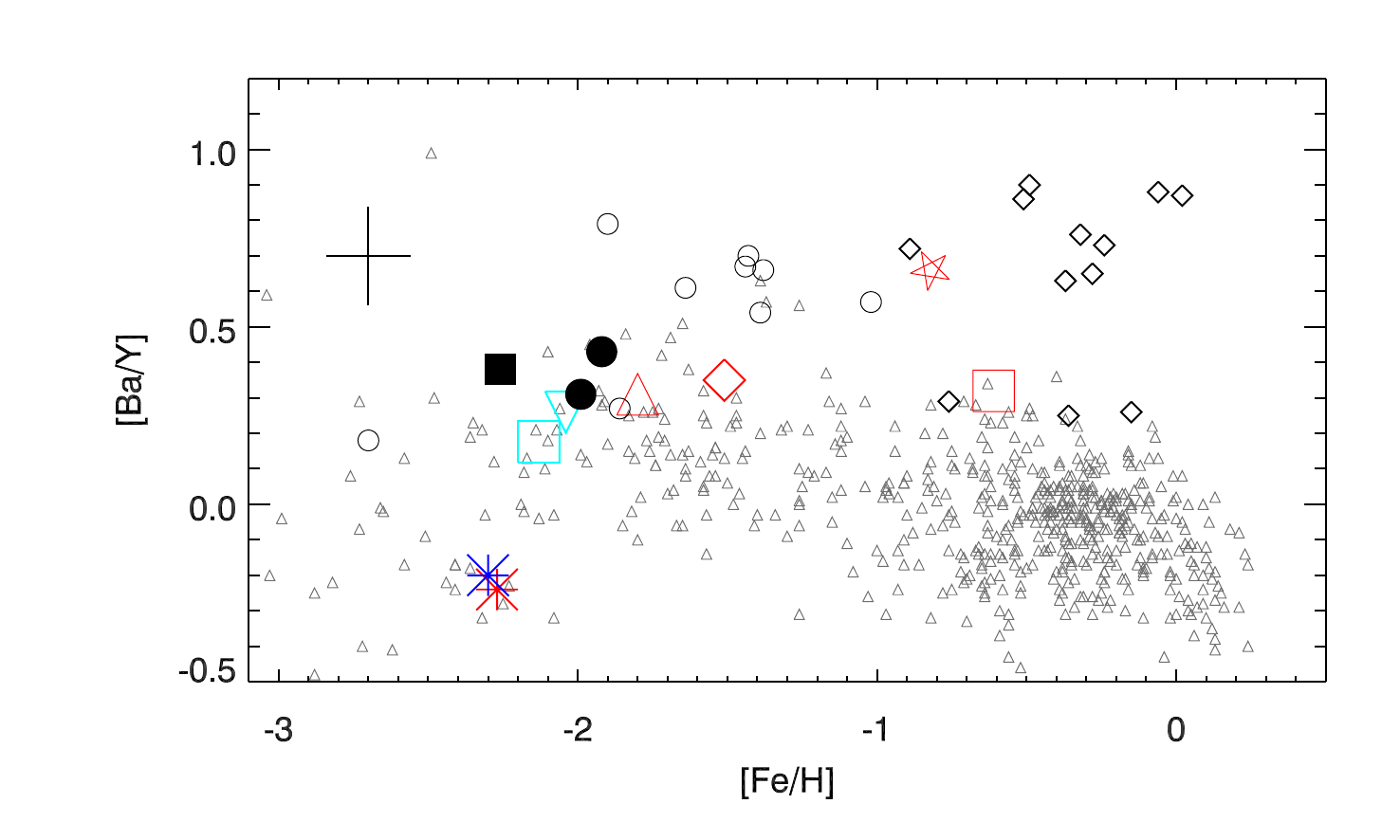}
      \caption{[Ba/Y] plotted against [Fe/H], see Fig. \ref{y_fe_fig} for the symbol legend.}
         \label{bay_fe_fig}
   \end{figure}

\begin{figure}
   \centering
   \includegraphics[width=\hsize]{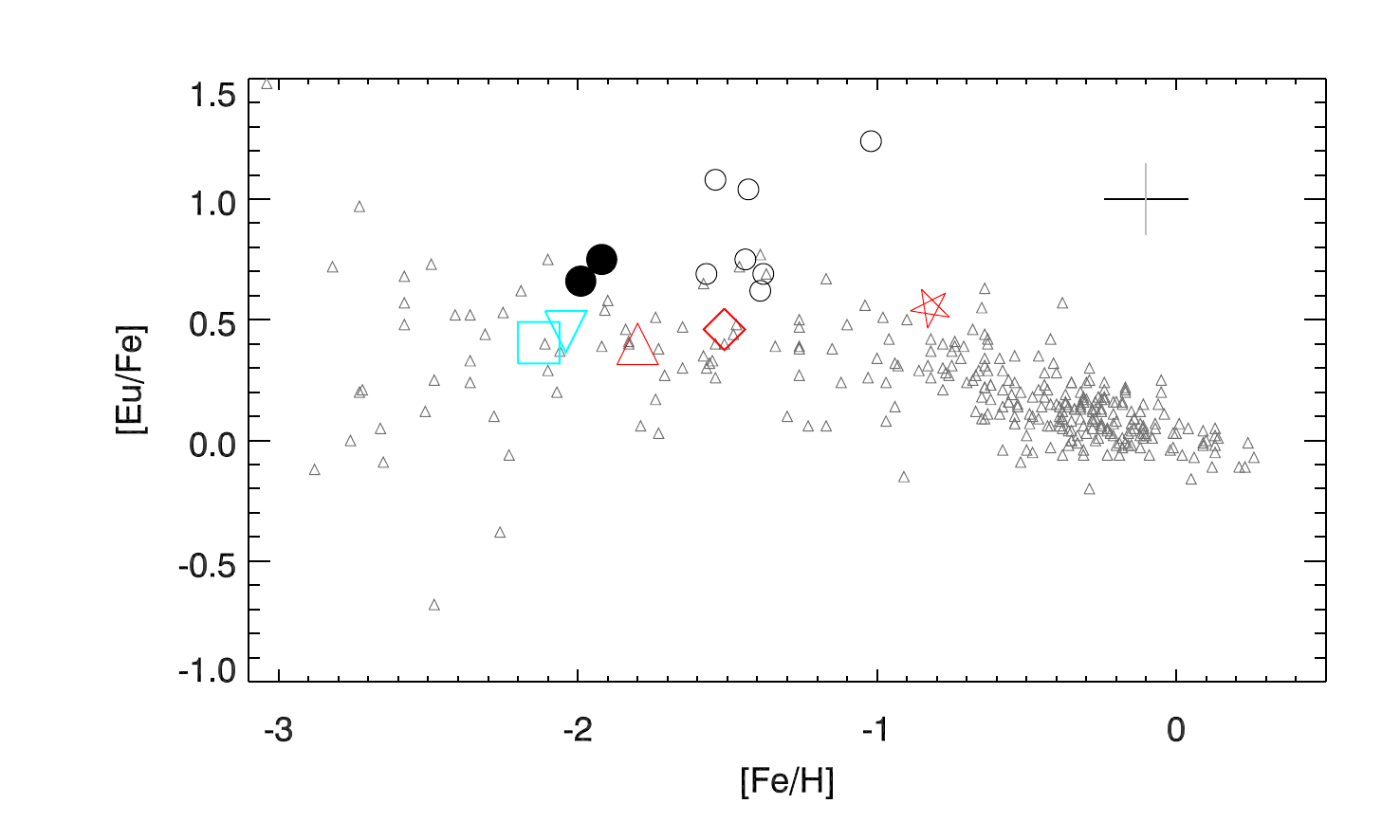}
      \caption{[Eu/Fe] plotted against [Fe/H], see Fig. \ref{alpha_fe_fig} for the symbol legend.}
         \label{eu_fe_fig}
   \end{figure}

{  The $\alpha$ elements Mg, Si, and Ca  are
all enhanced with respect to iron by $\sim 0.3$ to 0.5\,dex.
 Titanium is also enhanced by about 0.3 dex; one should 
note, however,  that  nucleosynthetically it  is not a pure $\alpha$ element,
since it may be synthesized  in nuclear statistical equilibrium,
together with iron-peak elements.
The odd light element Al is only detected in star
NGC 5634-3 and it is strongly enhanced over iron (0.6\,dex).
If on the one hand this is not too surprising, since this star
is also enhanced in Na, it is  surprising that this does not appear
to be accompanied by a decrease in Mg abundance as is usually
observed \citep[see e.g.][]{g01}.
The iron peak elements, V, Cr, Co, and Ni (Fig. \ref{ni_fe_fig}) follow the iron abundance,
while Sc seems to be slightly 
{\bf enhanced with respect to} iron in both stars,
and Mn and Cu slightly underabundant with respect to iron. 
In fact Mn is found to be  underabundant both in Sgr dSph stars
\citep{McW03,McW03b,sbordone07} and in Milky Way
stars \citep{g89,C04} at low metallicity. \citet{berg08}
presented NLTE computations that increase the Mn abundances
so that that [Mn/Fe] $\sim 0$.
Copper is measured in NGC 5634-3 only, and by a single line.
The fact that is underabundant with respect to iron is 
coherent with what is observed in Galactic stars
\citep{Mishenina,Bihain}.
One should note,  however,  that \citet{Bonifacio10}
warned against the possible effects of granulation
and NLTE that may affect the determination of Cu 
abundances, based on the differences found
between the abundances in dwarfs and giants in
the Globular Cluster NGC\,6397, which has a metallicity
similar to NGC\,5634. 
Zinc is slightly underabundant with respect to iron
in both stars, not inconsistent with the Galactic
trend at this metallicity \citep{Mishenina,Bihain},
but also not inconsistent with what is observed in Sgr dSph stars
at higher metallicity \citep{sbordone07}. 
The neutron capture element Y (Fig. \ref{y_fe_fig}) is underabundant with respect
to iron, consistent with what is observed in the Globular
Cluster NGC\,6397 \citep{James,lind11} and in the field stars
of similar metallicity \citep{Burris,Fulbright,Mashonkina01}, but
again, not inconsistent with what is observed in the higher metallicity 
stars of Sgr dSph \citep{sbordone07}.
The average [Ba/Fe] of   the two stars is nearly zero (Fig. \ref{ba_fe_fig}), which
is at variance with what is observed in  NGC\,6397 \citep{James,lind11},
where [Ba/Fe] is $\sim -0.2$, although it should be stressed
that in Galactic stars we see a large scatter in [Ba/Fe]
at this metallicity \citep{Burris,Fulbright,Mashonkina03}.
Europium is strongly enhanced over iron in both stars,
even more than what is found in NGC\,6397 \citep{James,lind11}, and we  also observe a large scatter in [Eu/Fe]
among Galactic stars at this metallicity  \citep{Burris,Fulbright,Mashonkina03},
while it has been found essentially at the solar value in the two
solar metallicity  Sgr dSph stars analyzed by \citet{Bonifacio00}.
}

 \subsection{\nci}
\label{results_5053}

The single star analyzed in \nci\ shows a metallicity of [\feii/H]=$-2.26\pm$0.01, in excellent agreement with the value of $-$2.27 in \citet{harris96}, based in large part on the Ca triplet analysis of 11 stars by \citet{geisler95}. 

The choice of employing \feii\ as the reference iron abundance and of assuming photometric atmosphere parameters has relevant effects on the [X/Fe] abundance ratios listed in Tab. \ref{abund_table}; since [\fei/H] is 0.22 dex lower than [\feii/H], if [\fei/H] were used as reference, as is usually done for neutral species, all the neutral [X/Fe] ratios would be 0.22 dex {\em higher}. The choice of using \feii\ as reference was motivated by the belief that \fei\ is significantly affected by NLTE and 3D effects in this star. However, at the present time NLTE / 3D corrections are only available  for a handful of species, so that we generally do not know whether the same is true for other neutral species, and to what extent. If we assumed, for instance, that a given neutral species was affected in the same way as \fei, for that species the ratio against \fei\ would be the appropriate one, while the ratio against \feii\ is the correct one if one assumes the species to be unaffected by NLTE/3D. As such, star \ncis\ is difficult to plot in Figs. \ref{alpha_fe_fig} and \ref{ni_fe_fig}: to  make the issue visible, we plotted the results for this star with a double symbol, the filled one indicating the abundance ratio as listed in Table \ref{abund_table}, the open one plotting instead [X/\fei] vs. [\fei/H]. 

Sodium is measured in \ncis\ only through the 568.8nm line and delivers a fairly high abundance ([Na/\feii]=0.21), about 0.4 dex higher than expected for a first-generation star. The NLTE correction computed through {\tt INSPECT} for this line (assuming \glog=1.0) is of -0.04 dex (in the sense of the NLTE-corrected abundance being lower). This would hint at the presence of multiple stellar generations in \nci\ as well.

{
While the present paper was undergoing the refereeing process,
an independent analysis of NGC\,5053 based on medium resolution
spectra (R$\sim 13\,000$)
appeared as a preprint \citep{boberg}.
They observed star NGC5053-69, (star  6 \relax in their list).
In spite of the lower resolution and limited spectral range, their derived 
atmospheric parameters are in remarkably good agreement with ours
(130 K difference in Teff, -0.05 \relax in log g and -0.07\,dex in [Fe/H].
They manage to measure [O/Fe] from the [OI] 630\,nm line and find [O/Fe]=-0.2,
which is consistent with their enhanced sodium ([Na/Fe]=0.6). Sodium is one
of the  most discrepant elements with respect to our analysis. We did not use
the same lines:   they used the 616.2\,nm line, while we used the 568.2\,nm
and 568.8\,nm lines. We also adopted different NLTE corrections; 
\citet{boberg} used those of \citet{gratton99}. 
The other element that is distinctly different between the two analyses
is Ca. \citet{boberg} derive [Ca/Fe]=+0.49, while we derive +0.10.
Considering that our Ca abundance relies on nine \ion{Ca}{i} lines and 
[Ca/Fe] is in excellent agreement with [Si/Fe] in this star (also derived
from nine lines), we believe that our result is very robust.
Given that no details on the lines used are given in the preprint, we cannot
make a hypothesis about the reason of this discrepancy.
With the two exceptions of Na and Ca, the agreement on the abundance ratios
for the other elements in common (Ti, Ni, Ba) is excellent.

Strong underabundances of Cr and Mn, with respect to iron,
are observed in this star, in line with what is observed
in Milky Way stars; in both cases this is likely
due to the neglect of NLTE effects \citep{Bonifacio09,BC10,berg08}.

Both the neutron capture elements Y and Ba are found to be underabundant
with respect to iron. For Y this is similar to what
is observed in the more metal-rich stars of Sgr dSph \citep{sbordone07}.
For Ba instead, all the giant stars observed by \citep{sbordone07}
are enhanced in Ba. Again, we would like to stress
that for both elements a large scatter is observed
in Galactic stars at these metallicities \citep{Burris,Fulbright,Mashonkina01,Mashonkina03}.
}

\subsection{{ Assessing the proposed association with Sgr dSph}}

The association of \nci\ and \ncs\ with Sgr dSph \citep{bellazzini03,law10} has  so far been exclusively based on their position and kinematic properties, and is thus probabilistic. One of the obvious reasons for exploring the chemistry of these clusters is thus to look for characteristics that may set them aside from the typical MW behavior at similar metallicities, but that are shared with known Sgr dSph populations. Such a  chemical signature is rather dramatic at higher metallicity \citep{cohen04,sbordone07}, but it becomes more and more difficult to discern as metallicity decreases, since Sgr dSph chemical evolution at low metallicities appears to closely match the values observed in the MW halo \citep[e.g.][]{mcwilliam13}. { We will now look in  greater  detail at some chemical markers that might be used to infer an association of \nci\ and \ncs\ with Sgr dSph.}

In Fig. \ref{alpha_fe_fig}, { the [X/Fe] ratios for the three $\alpha$-elements Mg, Si, and Ca, and the mixed-$\alpha$-Fe-peak Ti in \nci\ and \ncs\ are compared } with Sgr dSph main body populations, the MW halo, globular clusters associated with Sgr dSph, and globular clusters thought to have originally formed in the MW. Since  the \alphafe\ ratio in Sgr dSph closely follows the MW value below [Fe/H]$\sim-$1.2, we do not expect to see any odd behavior here. In fact, \ncs\ behaves as expected. The one star analyzed in \nci,\, on the other hand, displays strikingly low { [Ca/Fe] and [Ti/Fe]} when \feii\ is employed as reference, but it { would} fall in line with all the other displayed populations if \fei\ were used as reference. 

{Nickel} (Fig. \ref{ni_fe_fig}) is another element whose ratio to iron is characteristically low in metal-rich Sgr dSph populations with respect to the MW. However, the Galactic distribution of [Ni/Fe] becomes more dispersed at lower metallicity. While \nci\ and \ncs, together with confirmed Sgr dSph system member clusters, remain toward the lower end of the MW field stars [Ni/Fe] values, they are fully compatible with the abundances in \object{NGC 5897} and \object{NGC 6397}. 

      \begin{figure}
   \centering
   \includegraphics[width=\hsize]{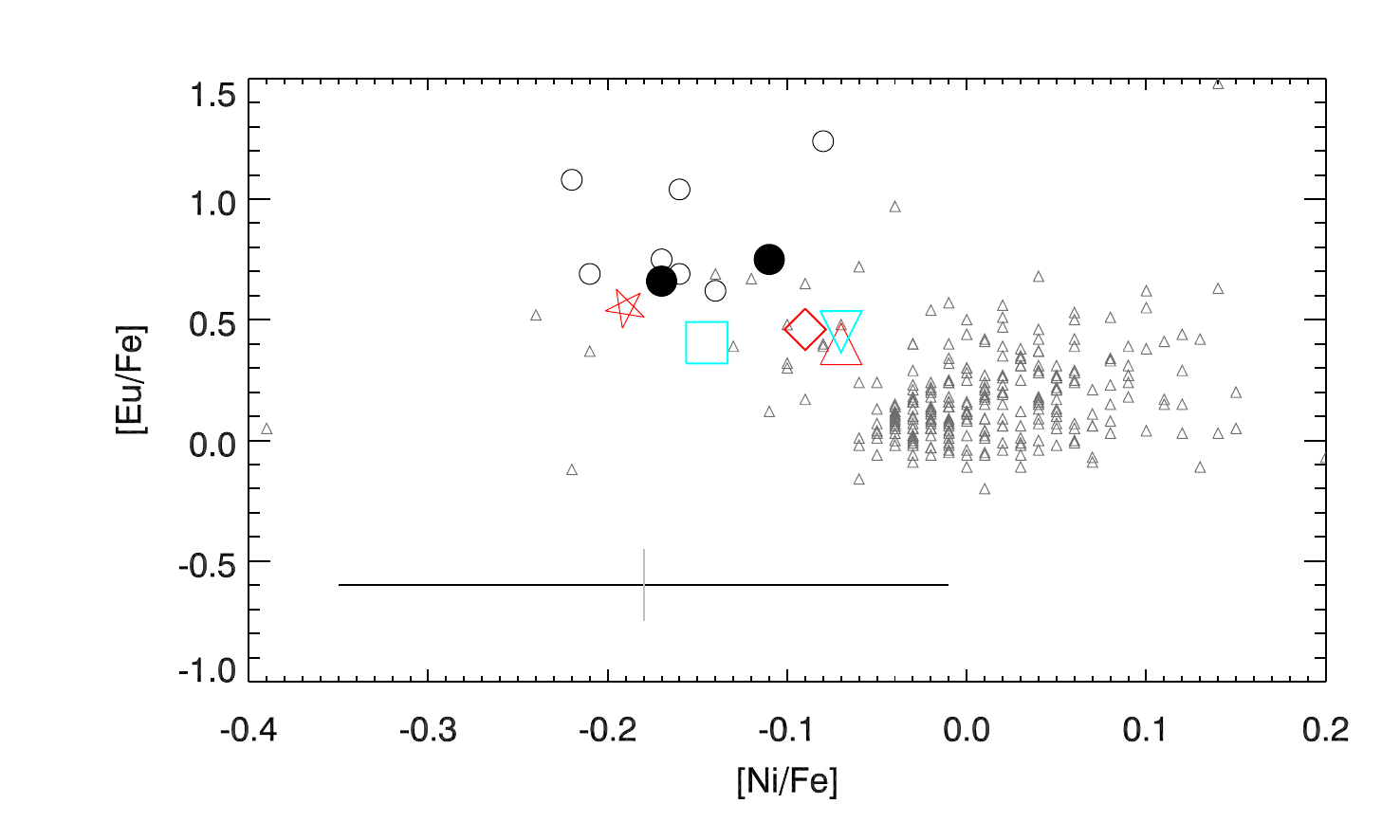}
      \caption{[Eu/Fe] plotted against [Ni/Fe], see Fig. \ref{alpha_fe_fig} for the symbol legend.}
         \label{nieu_fig}
   \end{figure}

Yttrium { and  barium abundances (see Figs. \ref{y_fe_fig} and \ref{ba_fe_fig}) as well as their ratio (Fig. \ref{bay_fe_fig})} are also rather peculiar in the more metal-rich Sgr dSph populations \citep[see e.g. Fig. 12 in][]{carretta14}. Yttrium in \ncs\ and, in particular, in \nci, appears  lower than in the MW halo population, and is coherent with both metal-rich and metal-poor Sgr dSph main body populations. Both \nci\ and \ncs\  actually appear a little more yttrium-poor than the other known Sgr dSph clusters, with the exception of \object{Palomar 12}. A somewhat puzzling case here is \object{Terzan 8} whose [Y/Fe] ratio is derived in \citet{mottini08} and \citet{carretta14}, but is reported on average as 0.4 dex lower in the former. The difference persists in the one star the two studies have in common where Y was measured. Since the \citet{mottini08} abundance is in better agreement with the general trend of Sgr dSph populations, we  plot it as well (as a blue large asterisk) in Figs. \ref{y_fe_fig} and \ref{bay_fe_fig}.  At any rate, again, \object{NGC 5897} and \object{NGC 6397} are not significantly removed from what is found in \ncs. 

{ Barium is generally enhanced in the more metal-rich Sgr dSph populations, but it has been known for a while that a subpopulation exists of Ba-poor Sgr dSph stars \citep[Fig. 5 in ][and Fig. \ref{ba_fe_fig} here]{sbordone07}, whose origin is unclear. The high [Ba/Fe] ratio, and the connected high [Ba/Y] ratio, are among the Sgr dSph chemical signatures that persist to the lowest metallicity, as shown in Fig. \ref{bay_fe_fig}. However, even this chemical peculiarity disappears around [Fe/H]=-2.0, most likely as an effect of the diminishing relevance of s-process at low metallicities. Seen in this context, \nci\ and \ncs\ nicely follow a general trend toward low [Ba/Fe] at low metallicity, and higher-than-MW values at high metallicity, which is discernible among the Sgr dSph core populations (with the notable exception of the aforementioned low-Ba population), and  the associated globular clusters. However, as said above, \nci\ and \ncs\ Ba abundances are also  compatible with MW values, and a similar picture is also drawn  by the [Ba/Y] ratio.}

    \begin{figure}
   \centering
   \includegraphics[width=\hsize]{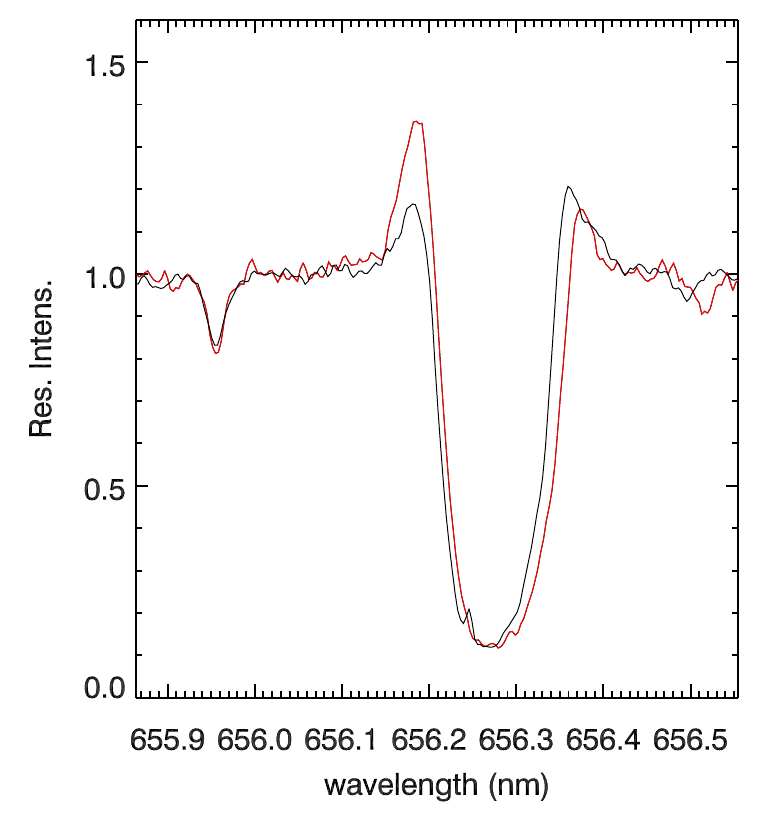}
      \caption{{\bf The H$\alpha$ line in the spectra of the} stars \ncsd\ (black) and \ncst\ (red), normalized and shifted to zero velocity. The absorption line on the blue side is the \ion{Ti}{ii} 655.959 nm.}
         \label{5634_halpha_fig}
   \end{figure}

Finally, [Eu/Fe] { in \ncs\ } appears to be somewhat higher than typical for MW stars of comparable metallicity and for \object{NGC 5897} and \object{NGC 6397}. Europium also shows a higher abundance here than in any other Sgr dSph cluster with the exception of the much more metal-rich \object{Pal 12}.

It is interesting to couple the results for Ni and Eu (as we do in Fig. \ref{nieu_fig}). Here, the contemporary low Ni and high Eu abundance observed both in the Sgr dSph metal-poor population and in \ncs\ sets them apart from the MW field and, to a lesser degree, from \object{NGC 5897} and \object{NGC 6397}. Among Sgr dSph clusters, the same locus is shared by \object{Pal 12} only, which is much more metal-rich than \ncs, while more metal-poor Sgr dSph clusters appear to agree more with the abundances found in MW globulars. { Although this result is intriguing, its significance is limited by the rather large error bar on Ni abundances.}

\section{Conclusions}

We present detailed chemical abundance studies of the distant halo globular clusters \nci\ and \ncs\  based on three luminous giant stars. The star \ncis\ has a metallicity of [\feii/H]=$-2.26\pm$0.10, while the two stars analyzed in \ncs\ have metallicities of [\fei/H]=$-1.99\pm$0.07 and $-1.92\pm$0.08 for \ncsd\ and \ncst, respectively (uncertainties representing internal line-to-line scattering).

Star \ncsd\ appears to be O-rich and Na-poor, while \ncst\ is 0.4 dex richer in Na, and 0.3 dex poorer in O, thus indicating the presence in \ncs\ of the Na/O anticorrelation, almost universally observed in globular clusters. The high Na abundance in star \ncis\ is also not consistent with the value expected in a first-generation star, { indicating -- in agreement with \citet{boberg} --} that \nci\ hosts multiple stellar populations as well. 

On the basis of their kinematics, both clusters are strongly suspected to have formed in the Sagittarius dwarf spheroidal galaxy, and to have been subsequently stripped by tidal interaction with the Milky Way. Although the Sgr dSph has a very characteristic set of chemical abundances in its metal-rich population, its more metal-poor stars resemble closely a typical halo composition, making it hard to use chemical abundances to assess the origin of these two clusters in the Sgr dSph system. Hints in this sense exist: \nci\ has a remarkably low yttrium abundance, \ncs\ a higher than usual europium content, a finding which, when coupled with its nickel abundance, makes it strikingly similar to metal-poor Sgr dSph populations, and more different from MW field and GC stars. More generally, both clusters appear to fall, chemically, closer to Sgr dSph populations than to halo populations. 
{ All these conclusions combined make an origin for both clusters (\ncs\ in particular) in the Sgr dSph system an appealing possibility. However, none of them appears strong enough to firmly confirm or reject this attribution.}


\begin{acknowledgements}
Support for L. S. and S. D. is provided by Chile's Ministry of Economy, Development, and Tourism's Millennium Science Initiative through grant IC120009, awarded to The Millennium Institute of Astrophysics, MAS. 
C.M.B. gratefully acknowledges the support provided by Fondecyt reg. n.1150060. 
S.V. gratefully acknowledges the support provided by Fondecyt reg. n. 1130721. 
M. B. acknowledges financial support from PRIN MIUR 2010-2011, project ``The Chemical and Dynamical Evolution of the Milky Way and Local Group Galaxies'', prot. 2010LY5N2T.
D.G. gratefully acknowledges support from the Chilean BASAL  Centro de Excelencia en Astrof\'isica y Tecnolog\'ias Afines (CATA) grant PFB-06/2007. 
E.C. is grateful to the FONDATION MERAC for funding her fellowship. 
Based in part on data collected at Subaru Telescope and obtained from the SMOKA, which is operated by the Astronomy Data Center, National Astronomical Observatory of Japan. This research has made use of NASA's Astrophysics Data System, and of the VizieR catalogue access tool, CDS, Strasbourg, France
\end{acknowledgements}


\Online

\begin{appendix} 
\section{Line-by-line \mygi\ fits}
\label{mygi_appendix}

\mygi\ operates by fitting a region around each relevant spectral line against a grid of synthetic spectra, and deriving the best-fitting abundance directly, rather than using the line equivalent width (EW), which {\bf \mygi\ computes, but mainly to estimate \Vturb}. While this makes \mygi\ arguably more robust against the effect of line blends, it makes it rather pointless to provide a line-by-line table of EWs, atomic data, and derived abundances,{ unless atomic data for all the features included in the fitted region are provided, which is quite cumbersome.}

\begin{figure}
   \centering
   \includegraphics[width=\hsize]{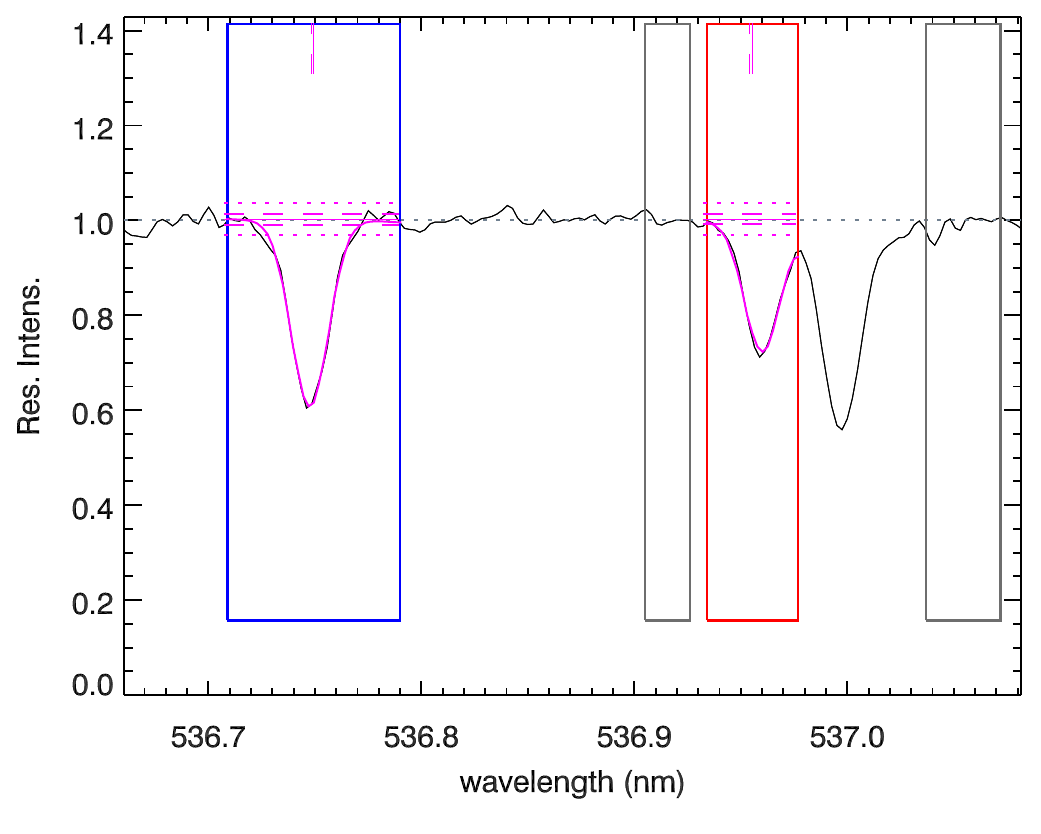}
      \caption{An example of \mygi\ fit for two features of star \object{NGC5634-2}. The blue box corresponds to a \fei\ feature, while the red box is a \ion{Co}{i} feature. Gray boxes are pseudo-continuum estimation intervals. Observed pseudo-normalized spectrum is in black; magenta profiles are best-fitting profiles for each region. The black dotted horizontal line is the pseudo-continuum level, while the continuous thin magenta horizontal line represents the best-fit continuum for the feature. Around it, dashed and dotted horizontal magenta lines represent 1-$\sigma$ and 3-$\sigma$ intervals of the local noise (S/N=88 in this area). Vertical dashed and continuous lines mark the theoretical feature center, and the actual center after the best-fit, per-feature Doppler shift has been applied.
              }
         \label{mygi_fit_figure}
   \end{figure}

{ To allow verification and comparison of our results, we thus decided to provide the per-feature abundances as well as the actual observed and synthetic best-fitting profile for each fitted region. We describe here in detail how these results will be delivered, since we plan to maintain the same format for any future \mygi-based analysis. 

The ultimate purpose of these comparisons is to ascertain whether abundance differences between works stem from any aspect of the line modeling, and assess the amount of the effect. We believe that providing the observed and synthetic profiles is particularly effective in this sense for a number of reasons:
\begin{itemize}
\item The reader can apply whatever abundance analysis technique he/she prefers to either the observed or the synthetic spectra. For instance, if   EWs are employed, he/she can determine the EW of either profile, and derive the abundance with his/her choice of atomic data, atmosphere models, and so on. This would allow him/her to directly determine what abundance the chosen method would assign to our best-fitting synthetic, and to compare it with the one we derive. 
\item The reader can evaluate whether our choice of continuum placement corresponds or differs from the one he/she applied, and assess the  broadening we applied to the syntheses, both looking at the feature of interest, and looking for any other useful feature  (e.g. unblended \fei\ lines, which are often used to set the broadening for lines needing synthesis).
\item The reader can directly assess the goodness of our fit with whatever estimator he/she prefers.
\item The reader gains access to the actual observed data we employed for every spectral region we used. Not every spectrum is available in public archives, and even fewer are available in reduced, and (possibly) coadded form. 
\end{itemize}

An example of \mygi\ fit results is presented in Fig. \ref{mygi_fit_figure}. To make the results easy to handle by Vizier, they were split into two tables:

\begin{itemize}

\item The features table contains basic data for each region successfully fitted in every star. This includes a code identifying the star (e.g. {\tt N5053\_69}), the ion the feature was used to measure, the starting and ending wavelength, the derived abundance, the EWs for the observed and best-fitting synthetic (determined by integration under the pseudo-continuum), the local S/N ratio, the small Doppler and continuum shift that MyGIsFOS allows in the fitting of each feature, flags to identify the \fei\ features that were used in the \Teff\ and \Vturb\ fitting process (the flags are not relevant for features not measuring \fei), and finally a feature code formed of the element, ion, and central wavelength of the fitted range (e.g. {\tt 3000\_481047}). This last code is unique to each feature, and it can be used to retrieve the profile of the fit from the next table.

\item The fits table contains all the fitted profiles for all the features for all the stars. Each line of the table corresponds to a specific pixel of a specific fit of a specific star, and contains the star identifying code, then the feature identifying code, followed by the wavelength, and the (pseudonormalized) synthetic and observed flux in that pixel. The user interested, for instance, in looking at the fit of the aforementioned \ion{Zn}{i} feature in star \ncis, has simply to select from the table all the lines beginning with ``{\tt N5053\_69 3000\_481047}''.

\end{itemize}

}
{ Both tables} are available through CDS. To reproduce the fit as plotted in Fig. \ref{mygi_fit_figure}, the synthetic flux must be {\em divided} by the continuum value provided { in the features table}.

\end{appendix}

\end{document}